\documentstyle[aps,preprint,tighten,eqsecnum,epsf]{revtex}
\newcommand{\myenn}{\boldmath \mbox{$n$}}
\newcommand{\myvee}{\boldmath \mbox{$v$}}
\newcommand{\mydoubleu}{\boldmath \mbox{$w$}}
\newcommand{\myEEE}{\boldmath \mbox{$E$}}
\newcommand{\myeta}{\boldmath \mbox{$\eta$}}
\newcommand{\myphi}{\boldmath \mbox{$\varphi$}}
\begin{document}
\preprint{TUW-95-21, gr-qc/9508028}
\title{On the Canonical Reduction of Spherically 
Symmetric Gravity\footnote{Revised version (to appear in
{\em Classical and Quantum Gravity}).}}
\author{Stephen R. Lau\footnote{email
address: lau@thp16.tuwien.ac.at}}
\address{Institut f\"{u}r Theoretische Physik\\
Technische Universit\"{a}t Wien\\
Wiedner Hauptstra\ss e 8-10\\
A-1040 Wien, AUSTRIA}
\maketitle
\begin{abstract}
In a thorough paper Kucha\v{r} has examined the
canonical reduction of the most general action 
functional describing the geometrodynamics of the 
maximally extended Schwarzschild geometry. This 
reduction yields the true degrees of freedom for 
(vacuum) spherically symmetric general relativity
(SSGR). The essential technical ingredient in 
Kucha\v{r}'s analysis is a canonical transformation 
to a certain chart on the gravitational phase
space which features the Schwarzschild mass parameter 
$M_{\scriptscriptstyle S}$, 
expressed in terms of what are essentially 
Arnowitt-Deser-Misner variables, as a canonical 
coordinate. (Kucha\v{r}'s paper complements earlier 
work by Kastrup and Thiemann, based mostly on Ashtekar
variables, which has also explicitly isolated the true 
degrees of freedom for vacuum SSGR.)
In this paper we discuss the geometric interpretation 
of Kucha\v{r}'s canonical transformation
in terms of the theory of quasilocal energy-momentum
in general relativity given by Brown and York. We find
Kucha\v{r}'s transformation to be a ``sphere-dependent 
boost to the rest frame," where the ``rest frame'' is 
defined by vanishing quasilocal momentum.
Furthermore, our formalism is general enough to
cover the case of (vacuum) two-dimensional dilaton 
gravity. Therefore, besides reviewing Kucha\v{r}'s 
original work for Schwarzschild black holes from the 
framework of hyperbolic geometry, we present new 
results concerning the canonical reduction of 
Witten-black-hole geometrodynamics. Finally, addressing 
a recent work of Louko and Whiting, we discuss some 
delicate points concerning the canonical reduction
of the ``thermodynamical action," which is of central 
importance in the path-integral formulation of 
gravitational thermodynamics.
\end{abstract}
\vskip 2mm
Vienna, February 1996
\newpage
\section{Introduction}
In a thorough paper
(Ref. \cite{Kuchar}, hereafter referred to as KVK)
Kucha\v{r} has examined the
canonical reduction of the most general action functional
describing the geometrodynamics of the maximally extended
Schwarzschild geometry. This reduction yields the true degrees
of freedom for vacuum general relativity subject to
ansatz of spherical symmetry. The key technical ingredient 
in Kucha\v{r}'s analysis is a canonical transformation to
a certain chart on the
gravitational phase space which features the Schwarzschild
mass parameter $M_{\scriptscriptstyle S}$, 
expressed in terms of what are essentially
Arnowitt-Deser-Misner (ADM) variables,\cite{ADM} as a 
canonical coordinate. (Kucha\v{r}'s paper 
complements earlier work\cite{Thiemann} by Kastrup and 
Thiemann, based mostly on Ashtekar variables, which has 
also explicitly isolated the true degrees of freedom for 
vacuum SSGR.) 
Potential applications of the new reduced
formalism include examinations, from the canonical viewpoint, 
of spherically symmetric collapse, the Hawking effect, and 
equilibrium gravitational thermodynamics. 
Indeed, Louko and Whiting have already made such an application. 
Applying Kucha\v{r}'s method to a spatially bounded exterior 
region of the Schwarzschild black hole, they have constructed 
the Schwarzschild thermodynamical (canonical) partition 
function completely within the Lorentzian Hamiltonian 
framework (Ref.\cite{LW}, hereafter referred to as LW).
Their canonical partition function is in agreement with
previous results derived via the Euclidean-path-integral
method.\cite{BMYW} The starting point in LW is the
``thermodynamical action,'' which is of central importance
in the path-integral formulation of gravitational 
thermodynamics. A very delicate issue in the analysis of LW 
concerns the treatment of the thermodynamical action's 
boundary terms under the canonical reduction via the 
KVK method.

In this paper we discuss the geometric interpretation of
Kucha\v{r}'s canonical transformation. By appealing
to notions of quasilocal energy and momentum in general
relativity which have been given by Brown and 
York\cite{BY,BYL}, we interpret Kucha\v{r}'s canonical 
transformation as a ``sphere-dependent boost to the rest 
frame," where the ``rest frame'' is defined by vanishing 
quasilocal momentum. It seems to us that KVK finds essentially 
the same interpretation via an alternate route involving 
subsequent canonical transformations and a ``parameterization 
at infinity.'' We believe that our approach complements the
one given by KVK. The center-stone of ours is
the following observation. On 
an arbitrary (spherically symmetric) spatial slice 
$\Sigma$ the parameter $\varphi$ describing the local boost
between the slice Eulerian observers at a point and the
rest-frame observers at the same point can be constructed
from the canonical variables of $\Sigma$ (if known in a tiny
spatial region surrounding the point of interest). 
Furthermore, we work in a framework which is general enough 
to cover Witten's two-dimensional dilaton gravity 
(2dDG).\cite{Witten} (Further still, many of our preliminary
results concerning the theory of quasilocal energy-momentum are 
unaffected by the inclusion of minimally coupled matter and 
hence are relevant for the two-dimensional dilaton-plus-matter 
model of Callan, Giddings, Harvey, and Strominger.\cite{CGHS})
Therefore, besides reviewing some of Kucha\v{r}'s original
work for Schwarzschild black holes from the framework of
hyperbolic geometry, we present new results concerning the
canonical reduction of Witten-black-hole geometrodynamics.
We show that the canonical transformation of KVK can also be 
made in the context of (vacuum) 2dDG. Therefore, the potential 
applications of the KVK 
formalism, listed in the first paragraph, are also relevant
for 2dDG.
Finally, with our general framework we address some of
the delicate points, first considered in LW, concerning the
canonical reduction of the ``thermodynamical action.''
Our conceptual framework supports the difficult technical
steps taken in LW. All of our results are given
for both the Schwarzschild and Witten-black-hole cases.

A few technical points demand some comment at the outset.
As mentioned, the analysis of KVK concerns the full Kruskal
spacetime, the maximally extended Schwarzschild geometry.
The canonical variables used in KVK are defined on spatial
slices which cut completely across the Kruskal diagram, and
therefore have to obey appropriate boundary conditions in the
asymptotic regions. In crossing from one spatial infinity to the
other, the slices of KVK are allowed to cross the horizon in
a completely general way. This introduces some technical
difficulties at the horizon, especially when one is considering
Kucha\v{r}'s canonical transformation. However, as demonstrated
in KVK, with care these difficulties may be surmounted.
We choose
to confine our attention entirely to the right static region of
the Kruskal diagram.  At first, we work with the time history of a
static region lying between concentric spheres. Thus we avoid
many of the technical difficulties faced by Kucha\v{r} at the
outset. We could, of course, work in the full Kruskal diagram,
but the essential points of this paper do not demand that we do
so. However, since we do chose to bypass a technical
treatment of the horizon, questions concerning how to handle such
horizon difficulties remain for the Witten black holes of 2dDG.
However, notationally we adopt nearly the same conventions as KVK.
Therefore, we expect that with the present paper as a stepping stone,
the interested reader could -with minimal effort- convert any and
all of the horizon arguments given in KVK into corresponding
arguments applicable to the Witten black-hole case.

The layout of this paper is as follows. In $\S$ {\sc II}, the preliminary
section, we describe the relevant kinematics of our spacetime
geometry. Since the spacetime geometry is spherically symmetric,
it proves convenient to work with a toy $1+1$ dimensional
spacetime $\cal M$. In reality, the points of $\cal M$ are round
spheres. In $\S$ {\sc III} we derive quasilocal\footnote{We
use the adjective ``quasilocal" because from
the four-dimensional viewpoint the energy and momentum
are associated round two-spheres. Thus these ``quasilocal"
quantities are associated with points of $\cal M$.} energy
and momentum expressions for the physical fields
defined on a generic spatial
slice $\Sigma$ of $\cal M$. The method used to derive the quasilocal
expressions is a Hamilton-Jacobi analysis of an appropriate action
principle for $\cal M$. In $\S$ {\sc IV} we use the developed notions of
quasilocal energy and momentum to underscore the geometric
significance of Kucha\v{r}'s canonical transformation. This
section also considers the reduction of the canonical action
with the boundary conditions adopted in this work.
Finally, $\S$ {\sc V} considers the canonical reduction of the
thermodynamical action.

\section{Preliminaries}

\subsection{Spacetime $\cal M$ and foliations}

Consider a $1+1$ dimensional spacetime region $\cal M$ which is
bounded spatially. The region $\cal M$ consists of a collection
of one dimensional spacelike slices $\Sigma$. The letter $\Sigma$
denotes both a foliation of $\cal M$ into spacelike slices and a
generic leaf of such a foliation. However, for the initial
spacelike slice we reserve the special symbol $t'$ (also the
value of the coordinate time on the initial slice), and,
likewise, for the final spacelike slice we reserve the symbol
$t''$ (also the value of the coordinate time on the final slice).
On spacetime $\cal M$ we have coordinates $(t, r)$,
and a generic spacetime point\footnote{One may consider each point
of $\cal M$ to be round two-sphere with radius $R(t,r)$, where $R$
is the radius function.} is $B(t,r)$. Every constant-$t$ slice $\Sigma$
has two boundary points $B_{i}$ (at $r = r_{i}$) and $B_{o}$
(at $r = r_{o}$). Assume that along $\Sigma$ the coordinate $r$
increases monotonically from $B_{i}$ to $B_{o}$. We represent the
timelike history $B_{i}(t) \equiv B(t, r_{i})$ by $\bar{\cal T}_{i}$
(unbarred $\cal T$ is reserved for another meaning) and refer to it
as the {\em inner boundary}. Likewise, we represent the timelike
history $B_{o}(t) \equiv B(t, r_{o})$ by $\bar{\cal T}_{o}$ and refer
to it as the {\em outer boundary}. Later on, when we deal with
black-hole solutions, we will ``seal" the inner boundary. In other
words, the time development at the inner boundary will be arrested,
and the point $B_{i}$ will correspond to a bifurcation point in a
Kruskal-like diagram. We denote the {\em corner} points of our
spacetime as follows: $B'_{i} \equiv B(t',r_{i})$, $B'_{o} \equiv
B(t',r_{o})$, $B''_{i} \equiv B(t'',r_{i})$, and $B''_{o} \equiv
B(t'',r_{o})$.

We can also consider a {\em radial} foliation of $\cal M$
by a family of one-dimensional timelike surfaces which extend
from $\bar{\cal T}_{i}$ outward to $\bar{\cal T}_{o}$ (for
black-hole solutions $\bar{\cal T}_{i}$
may be a degenerate sheet).\cite{Hayward}
These are constant-$r$ surfaces.
Like before, we loosely use the letter $\bar{\cal T}$ both to denote
the radial foliation and a generic leaf of this foliation. We
call $\bar{\cal T}$ leaves {\em sheets}, whereas we have called
$\Sigma$ leaves {\em slices} (this is an informal convention). 
Abusing the notation a bit, we also often let the symbol 
$\bar{\cal T}$ denote the total timelike boundary 
$\bar{\cal T}_{i} \bigcup \bar{\cal T}_{o}$. 
However, when the symbol $\bar{\cal T}$ has this meaning, it 
always appears in the phrase ``the $\bar{\cal T}$ boundary.''

\subsection{Metric decompositions}
The spacetime metric is $g_{ab}$. The metric on a generic
$\Sigma$ slice is $\Lambda^{2}$, and the metric on the
$\bar{\cal T}$ boundary is denoted by
$- \bar{N}^{2}$. In terms of the $\Sigma$ foliation, the
metric may be written in ADM form\cite{ADM}
\begin{equation}
g_{ab}{\rm d}x^{a} {\rm d}x^{b}
= - N^{2} {\rm d} t^{2} + \Lambda^{2}\left({\rm d}r + N^{r}
{\rm d}t\right)^{2}\, ,\label{ADM}
\end{equation}
with $N$ and $N^{r}$ denoting the familiar {\em lapse}
and {\em shift}. The vector field
\begin{equation}
u^{a} \partial/\partial x^{a} 
= N^{-1}\left(\partial/\partial t 
- N^{r} \partial/\partial r \right)
\end{equation}
is the unit, timelike, future-pointing normal to the 
$\Sigma$ foliation.

In terms of the $\bar{\cal T}$ foliation, the $\cal M$ metric 
takes the form
\begin{equation}
g_{ab}{\rm d}x^{a} {\rm d}x^{b}
=  \bar{\Lambda}^{2} {\rm d} r^{2}
- \bar{N}^{2}\left({\rm d}t +
\bar{\Lambda}^{t}{\rm d}r\right)^{2}\, ,
\end{equation}
where $\bar{\Lambda}$ and 
$\bar{\Lambda}^{t}$ are the {\em radial
lapse} and {\em radial shift}. 
The unit, spacelike, $\bar{\cal T}$-foliation
normal is
\begin{equation}
\bar{{{\myenn}}}^{a} \partial/\partial x^{a} = 
\bar{\Lambda}^{-1}\left(\partial/\partial r
- \bar{\Lambda}^{t}\partial/\partial t\right)\, .
\end{equation}
We also define the $\bar{\cal T}$ {\em boundary}
normal $\bar{n}^{a}$ on $\bar{\cal T}_{i}$ and $\bar{\cal T}_{o}$
by the requirement that it always be {\em outward}-pointing on
these boundary elements. On the outer boundary
$\bar{\cal T}_{o}$ the outward normal is
$\bar{n}^{a} = \bar{{{\myenn}}}^{a}$, while on the inner boundary
$\bar{\cal T}_{i}$ the outward normal is  
$\bar{n}^{a} = - \bar{{{\myenn}}}^{a}$. (For a handful
$\{n^{a}, \bar{n}^{a}, v, w, \eta, \varphi, E\}$ of symbols
we use this convention throughout the paper. Regular letters 
represent objects associated with the boundary and have the 
appropriate sign for each boundary element built in. 
Boldface versions of the same letters represent the same objects 
but with the fixed sign appropriate for the outer boundary. 
Perhaps it would have been a more natural choice to
adopt the reciprocal notation and let the boldface letters 
possess the sign flexibility. We have made the seemingly unnatural 
choice simply because it leads to the minimal use of boldface 
letters. In other words, in what follows the sign-flexible 
quantities happen to appear more frequently than their fixed-sign 
counterparts.)

By equating the coefficients of the above forms of $g_{ab}$, we obtain
the following relations between the ``barred" and ``unbarred" variables:
\begin{eqnarray}
\bar{N} & = & N/\gamma \label{transformation}
\eqnum{\ref{transformation}a} \\
\bar{\Lambda} & = & \gamma \Lambda
\eqnum{\ref{transformation}b}\\
\Lambda N^{r}/N & = & - \bar{N}
\bar{\Lambda}^{t}/\bar{\Lambda}\, .
\eqnum{\ref{transformation}c}
\addtocounter{equation}{1}
\end{eqnarray}
Here $\gamma \equiv (1 - {{\myvee}}^{2})^{-1/2}$ is the
local relativistic factor
associated with the velocity ${{\myvee}} \equiv \Lambda N^{r}/ N
= - \bar{N} \bar{\Lambda}^{t}/\bar{\Lambda} 
= - \bar{{{\myvee}}}$.\cite{Hayward} The timelike normal 
associated with the foliations 
$B_{i}(t)$ and $B_{o}(t)$ of the boundary sheets
$\bar{\cal T}_{i}$ and $\bar{\cal T}_{o}$ is 
$\bar{u}^{a} \partial /\partial x^{a} 
= \bar{N}^{-1} \partial/\partial t$. Note
that on the $\bar{\cal T}$ boundary the vector fields 
$u^{a}$ and $\bar{u}^{a}$ need not
coincide. Also, fixation of the $t$ coordinate gives a
collection of points $B(r)$ which foliates the slice $\Sigma$.
The normal associated with this foliation of $\Sigma$ is
${{\myenn}}^{a} \partial/\partial x^{a} 
= \Lambda^{-1}\partial /\partial r $. 
Again, we define a boundary
normal $n^{a}$ such that at the inner boundary
$n^{a}$ is $- {{\myenn}}^{a}$, the outward-pointing 
normal of $B_{i}$ as embedded in
$\Sigma$, while at the outer boundary 
$n^{a}$ is ${{\myenn}}^{a}$, the
outward-pointing normal of $B_{o}$ as embedded in $\Sigma$.
On the inner and outer boundaries $n^{a}$ and 
$\bar{n}^{a}$ need not coincide.
It is easy to verify the following point-wise boost relations:
\begin{eqnarray}
\bar{u}^{a} & = & \gamma u^{a} + v \gamma n^{a}
\label{boundaryboost} \eqnum{\ref{boundaryboost}a} \\
\bar{n}^{a} & = & \gamma n^{a} + v \gamma u^{a}\, ,
\eqnum{\ref{boundaryboost}b}
\addtocounter{equation}{1}
\end{eqnarray}
where $v = - {{\myvee}}$ on $\bar{\cal T}_{i}$ and
$v = {{\myvee}}$ on $\bar{\cal T}_{o}$. 

\subsection{Extrinsic curvatures}
The extrinsic curvature of a $\Sigma$ slice as embedded in
$\cal M$ is ${{\cal K}} \equiv - \nabla_{a}u^{a}$, where 
$\nabla_{a}$ denotes the torsion-free covariant derivative 
operator compatible with $g_{ab}$. 
One may also consider spacelike slices
$\bar{\Sigma}$ which are everywhere orthogonal by assumption
to the $\bar{\cal T}$ sheets. 
Since the spacetime-filling extension
of the $\bar{\cal T}$ sheets from $\bar{\cal T}_{i}$
to $\bar{\cal T}_{o}$ is arbitrary, the $\bar{\Sigma}$ slices
are almost as general as the $\Sigma$ slices. However, the
$\bar{\Sigma}$ slices are restricted by the requirement that
their normal vector field coincides with $\bar{u}^{a}$ on 
the $\bar{\cal T}$ boundary. We describe such slices as
{\em clamped}. When the velocity $v$
defined above is set to zero on
the boundary, then the $\Sigma$ slices are clamped. (In which
case, there is no longer a need to make a distinction
between barred and unbarred slices.) Also define 
$\bar{{{\cal K}}} \equiv - \nabla_{a}\bar{u}^{a}$,
the extrinsic curvature associated with the $\bar{\Sigma}$
slices.
The extrinsic curvature associated with the $\bar{\cal T}$
boundary elements is defined by $\bar{\vartheta} \equiv - 
\nabla_{a}\bar{n}^{a}$. We may also
consider a foliation $\cal T$ generated by the $u^{a}$ Eulerian
histories of points in the $\Sigma$ slices. At the boundary,
the $\cal T$ sheets may be ``crashing into" or ``emerging from"
the actual boundary elements $\bar{\cal T}_{i}$
and $\bar{\cal T}_{o}$. Nevertheless, one can define an
extrinsic curvature
$\vartheta \equiv - \nabla_{a} n^{a}$.
The value of $\vartheta$ at a particular boundary point of
the $\bar{\cal T}$ boundary is associated with the $\cal T$ sheet
intersecting that point. We have that $\vartheta = - n_{b}
a^{b}$, where $a^{b} \equiv u^{a}\nabla_{a} u^{b}$ is the
spacetime acceleration of $u^{a}$. Since by assumption the 
metric-compatible connection associated with $\cal M$ is 
torsion-free, $- \vartheta = n^{a} \nabla_{a} [\log N]$.

With our transformation equations (\ref{boundaryboost}),
one can derive the following ``splitting" formulae for
$\bar{{\cal K}}$ and $\bar{\vartheta}$:
\begin{eqnarray}
\bar{{{\cal K}}} & = & \gamma {{\cal K}} + v \gamma \vartheta -
\bar{n}^{a} \nabla_{a} \eta  \label{splittings}
\eqnum{\ref{splittings}a} \\
& &\nonumber \\
\bar{\vartheta} & = & \gamma \vartheta + v \gamma {{\cal K}} -
\bar{u}^{a} \nabla_{a} \eta\, , \eqnum{\ref{splittings}b}
\addtocounter{equation}{1}
\end{eqnarray}
where $\eta \equiv \tanh^{-1} v
= \frac{1}{2}\log |(1 + v)/(1 - v)|$.
Note that $\eta = \tanh^{-1} - {{\myvee}}$
on $\bar{\cal T}_{i}$ and $\eta = \tanh^{-1}
{{\myvee}}$ on $\bar{\cal T}_{o}$. In accord with our conventions
also set ${{\myeta}} \equiv \tanh^{-1} {{\myvee}}$.

\section{Action and quasilocal energy-momentum}
We begin this section by precisely defining the type
of action functional associated with our bounded
spacetime region $\cal M$ which is of interest in
this work. We will discuss in detail the action's
associated variational principle, paying strict
attention to all boundary terms. This is the
background work necessary to derive expressions for
the quasilocal energy and momentum
associated with a generic bounded slice $\Sigma$.

\subsection{Variational principle}
Our analysis begins with the following action functional:
\begin{eqnarray}
{\cal S}^{\it 1} & = & 
{\textstyle \frac{1}{4}}\,\alpha\!\int_{\cal M}
{\rm d}^{2}x \sqrt{-g} e^{- 2 \Phi}\left[ {\cal R} +
2y g^{ab} \nabla_{a}\Phi \nabla_{b}\Phi
+ 2\lambda^{2}y \exp\left[(2 - y)2\Phi\right]\right]
\nonumber \\
& &
+ {\textstyle \frac{1}{2}}\,\alpha\!\int^{t'}_{t''}
{\rm d}r \Lambda e^{- 2 \Phi}{{\cal K}}
- {\textstyle \frac{1}{2}}\,\alpha\!\int_{\bar{\cal T}}
{\rm d}t \bar{N} e^{- 2 \Phi} \bar{\vartheta}
- \left. {\textstyle \frac{1}{2}}\,\alpha e^{- 2 \Phi} \eta
\right|^{B''}_{B'}
\label{baseaction}\, ,
\end{eqnarray}
where ${\cal R}$ is the scalar curvature of $\cal M$ built
from the metric $g_{ab}$, and the scalar field $\Phi$ is the
celebrated {\em dilaton}. The variable $y$ is an as-yet
unspecified number, $\lambda$ is a positive
constant with dimensions of inverse length, and 
$\alpha$ is another positive (and possibly dimensionful)
constant. The ${\cal M}$ integral in our action corresponds
to a subclass of models within the larger framework
of generalized dilaton theories.\cite{Kummer}
In (\ref{baseaction}) and throughout the rest of this
paper we use the following short-hand notations:
\begin{eqnarray}
\int_{t'}^{t''} & = & \int_{t''} - \int_{t'}
\eqnum{\ref{Tshort}a} \\
\int_{\bar{\cal T}} & = & \int_{\bar{\cal T}_{i}} + 
\int_{\bar{\cal T}_{o}} \label{Tshort} 
\eqnum{\ref{Tshort}b} \\
\Bigl.\Bigr|^{B''}_{B'} & = &  
\Bigl.\Bigr|_{B''_{i}} - \Bigl.\Bigr|_{B'_{i}}
+ \Bigl.\Bigr|_{B''_{o}} - \Bigl.\Bigr|_{B'_{o}} 
\,\,\,\, ,
\eqnum{\ref{Tshort}c}
\addtocounter{equation}{1}
\end{eqnarray}
where (\ref{Tshort}a) is used only when
$t'$ and $t''$ represent the initial and final slices
and not integration parameters.
The boundary terms in the above action
ensure that its associated variational
principle features fixation of
induced metric and the dilaton on the
boundary. Symbolically, we
could collect the boundary terms into one expression
\begin{equation}
\left({\cal S}^{\it 1}\right)_{\partial {\cal M}} =
{\textstyle \frac{1}{2}}\,
\alpha\!\int_{\partial {\cal M}}{\rm d}x
{\textstyle \sqrt{|{}^{1}\! g|}} 
e^{-2\Phi} {{\cal K}}\, ,
\end{equation}
where here ${{\cal K}}$ is used for the extrinsic
curvature over the whole
boundary $\bar{\cal T}_{i} \bigcup \bar{\cal T}_{o} 
\bigcup t' \bigcup t''$.
The corner-point contributions in (\ref{baseaction}) 
are included because,
though the corner points are a set of measure zero 
in the integration of
$e^{-2\Phi} {{\cal K}}$ over all
of $\partial {\cal M}$, the term ${{\cal K}}$ becomes
infinite at the corners,
since the boundary normal changes discontinuously
from $u^{a}$ to $\bar{n}^{a}$ at
these points.\cite{Hayward}
Finally, we write ${\cal S}^{\it 1}$ for the action, because we
anticipate the need to append to the action a Gibbons-Hawking
subtraction term $- {\cal S}^{\it 0}[\bar{N},\Lambda,\Phi]$ (a
functional of the fixed boundary data).\cite{Gibbons,BY}
In this case the full action would be ${\cal S} \equiv
{\cal S}^{\it 1} - {\cal S}^{\it 0}$. We briefly 
consider the more general action
later in an appendix. Also, we could add to the
action a matter contribution ${\cal S}^{\it m}$. Most of our work
in this section on quasilocal energy-momentum would be unaffected 
by an ${\cal S}^{\it m}$ contribution to the action,
as long as the matter fields were minimally coupled. However,
an ${\cal S}^{\it m}$ 
contribution would affect the following sections
devoted to the canonical reduction. Therefore, for the sake
of simplicity we do not consider an ${\cal S}^{\it m}$ further.

We are interested in (vacuum) 
spherically symmetric general relativity
(SSGR). As was first noted by Thomi, Isaak, and 
H\'{a}j\'{\i}\v{c}ek,\cite{action} a suitable 
action for SSGR is given by (\ref{baseaction})
with the choices $y = 1$, $\alpha = \lambda^{-2}$. 
Note that for the SSGR case, we set $\alpha$
equal to the {\em dimensionful} constant $\lambda^{-2}$. This gives
our action the units of action in four dimensions.
An appendix shows that this action can be obtained via a
reduction by spherical symmetry of a covariant first-order 
version\cite{York,Hayward} 
of the Einstein-Hilbert action (where the {\em four dimensional} 
action principle is associated with the time history ${}^{4}\!{\cal M}$
of a spatial region bounded by concentric spheres).
In this correspondence with SSGR, it turns out that the radius of
a round sphere is given by
\begin{equation}
R = \sqrt{\frac{A}{4\pi}} =
\frac{e^{-\Phi}}{\lambda}\, ,\label{radius}
\end{equation}
where $A$ stands for the proper area of the sphere.

In this paper we also consider (vacuum) two-dimensional
dilaton gravity (2dDG), which corresponds to $y = 2$ and 
$\alpha$ taken to be a dimensionless positive number. 
Often $\alpha$ is chosen to be $2/\pi$, but $\alpha$ remains 
essentially arbitrary. 
The arbitrariness of $\alpha$ for vacuum 2dDG 
corresponds to the freedom to shift the dilaton 
by a constant $\Phi \rightarrow \Phi - \log \sqrt{\alpha'}$ 
(a freedom not present for the SSGR action). Under such a shift,
$\alpha \rightarrow \alpha'' = \alpha\alpha'$. We would actually
prefer to set $\alpha = 2$ for reasons which become clear
later. Adding the right matter contribution $S^{m}$ to the vacuum 2dDG
action, we would obtain the CGHS model.\cite{CGHS} Though we 
do not consider this matter contribution, the expressions for 
quasilocal energy-momentum that we derive in this section
are nevertheless valid for the full CGHS theory. 
For the SSGR case the action
${\cal S}^{\it 1}$ has dimensions of length-squared, while for
2dDG the action ${\cal S}^{\it 1}$ is dimensionless.
This difference in units will propagate throughout
all the formulae to follow. 
However, the freedom of allowing $\alpha$
to be either $\lambda^{-2}$ 
or a plain number will automatically
keep track of the correct units for both cases. For the SSGR
case  $(y = 1, \alpha = \lambda^{-2})$ all of our conventions
have been tailored to match those of KVK and LW.

The first step is to compute the variation of the action.
One can compute the variation in a number of ways, but the fastest
way is the following. Note that $\cal R$ is a pure divergence,
and then perform an integration by parts on the $\cal R$ term in the 
action.
This leads to cancelation of
most of the boundary terms. This short calculation and resulting
form of the action are given later in the discussion in the text
preceding (\ref{3+1action}). Vary the resulting form of the action
(\ref{3+1action}) to find
\begin{eqnarray}
\delta {\cal S}^{\it 1} & = &
\left({\rm terms\,\,giving\,\,the\,\,equations\,\,
 of\,\,motion}\right) \nonumber \\
& & + \int^{t''}_{t'}{\rm d}r
\left( P_{\Lambda}\delta\Lambda
+ P_{\Phi}\delta\Phi\right)
+ \int_{\bar{\cal T}}{\rm d}t
\left( \bar{\Pi}_{\bar{N}}\delta\bar{N}
+ \bar{\Pi}_{\Phi}\delta\Phi\right)
+ \left. \alpha
e^{- 2 \Phi} \eta \delta\Phi
\right|^{B''}_{B'}\, ,  \label{variation}
\end{eqnarray}
where we have defined the momenta
\begin{eqnarray}
P_{\Lambda} & \equiv & \alpha
e^{- 2\Phi} u^{a} \nabla_{a} \Phi \label{momenta}
\eqnum{\ref{momenta}a} \\
P_{\Phi} & \equiv & - \alpha
e^{- 2 \Phi} \Lambda
\left( {{\cal K}} + y u^{a} \nabla_{a} \Phi\right)
\eqnum{\ref{momenta}b} \\
\bar{\Pi}_{\bar{N}} & \equiv &
- \alpha e^{- 2 \Phi} \bar{n}^{a} \nabla_{a} \Phi
\eqnum{\ref{momenta}c} \\
\bar{\Pi}_{\Phi}  & \equiv & \alpha
e^{- 2 \Phi}\bar{N}
\left(\bar{\vartheta}
+ y \bar{n}^{a} \nabla_{a} \Phi\right)\, .
\eqnum{\ref{momenta}d}
\addtocounter{equation}{1}
\end{eqnarray}
Inspection of the
variation of the action shows that $P_{\Lambda}$
is the gravitational momentum conjugate to
$\Lambda$. Likewise, $\bar{\Pi}_{\bar{N}}$ is the
gravitational momentum conjugate to $\bar{N}$,
where now conjugacy is defined with respect
to the $\bar{\cal T}$ boundary. The momentum conjugate to
the dilaton field is $P_{\Phi}$. The momentum 
$\bar{\Pi}_{\Phi}$ is the $\bar{\cal T}$ boundary
analog of $P_{\Phi}$.
Note that one might try to define a momentum 
$p_{\Phi} \equiv \alpha e^{- 2 \Phi} \eta$ in some
sense conjugate to $\Phi$ at the corners.
However, we chose not to do this. Therefore,
strictly speaking, the canonical action we consider 
later has a mixed Hamiltonian-Lagrangian form. 
If the corner terms
in the action (\ref{baseaction}) had not been
included, then the variational principle would
have featured fixation of $\eta$ on the corner
points. Fixing $\Phi$ at the corners seems to be
more in harmony with the fact that $\Phi$ is
fixed on $t'$, $t''$, and the $\bar{\cal T}$ 
boundary.

Our momenta $P_{\Lambda}$ and $P_{\Phi}$ agree
with the analysis of KVK.
To see this take the SSGR case and use the
fact that $P_{R} = -(1/R) P_{\Phi}$. One then finds
precisely Kucha\v{r}'s momenta,
\begin{eqnarray}
P_{\Lambda} & = &
- N^{-1} R \left( \dot{R} - N^{r} R'\right)
\label{KVKmomenta} \eqnum{\ref{KVKmomenta}a}\\
P_{R} & = &
- N^{-1}\left[R\left(\dot{\Lambda}
- (\Lambda N^{r})' \right)
+ \Lambda\left(\dot{R} - N^{r} R'\right)\right]\, .
\eqnum{\ref{KVKmomenta}b}
\addtocounter{equation}{1}
\end{eqnarray}
To get the last expression, we have used the definition
of ${{\cal K}}$ given in the preliminary section to find
${{\cal K}} = - (N\Lambda)^{-1}\left[\dot{\Lambda}
- (\Lambda N^{r})'\right]$.
(Note, however, that due to well-entrenched notation
for the dilaton, we unfortunately must break with
the KVK convention of only using Greek
letters to represent spatial densities like
$\Lambda$. Though represented by a Greek letter, the
dilaton  $\Phi$ is a scalar and its momentum $P_{\Phi}$
is a density.)

\subsection{Quasilocal energy-momentum}
As advertised, our variational principle has been rigged
so that the induced metric
($\Lambda^{2}, -\bar{N}^{2}$) and the dilaton $\Phi$
are fixed as boundary data. In particular, the lapse of
proper time between the initial and final slices
is fixed in the variational principle, since this
information is encoded in the $\bar{\cal T}$ boundary metric
$- \bar{N}^{2}$. This is the key feature exploited
in the Brown-York theory\cite{BY,BYL}
of quasilocal energy-momentum in general
relativity. Following the basic line of reasoning in 
this theory, we ``read off" from the variation of the action
what geometric expressions play the role of quasilocal
energy and momentum in our theory.

We begin by writing the boundary terms in the
variation (\ref{variation}) of the action in the
following suggestive way:
\begin{equation}
\left(\delta {\cal S}^{\it 1}\right)_{\partial {\cal M}}
= - \int^{t''}_{t'}{\rm d}r
\left( J\delta\Lambda + \Lambda T \delta\Phi\right)
- \int_{\bar{\cal T}}{\rm d}t
\left(\bar{E}\delta\bar{N}
+ \bar{N} \bar{{S}} \delta\Phi\right)
+ \left. \alpha
e^{- 2 \Phi} \eta  \delta\Phi
\right|^{B''}_{B'}\, ,
\end{equation}
where we have defined the {\em scalars}
\begin{eqnarray}
\bar{E} & \equiv & - \bar{\Pi}_{\bar{N}}
= \alpha e^{-2 \Phi}
\bar{n}^{a} \nabla_{a} \Phi \label{qldensities}
\eqnum{\ref{qldensities}a} \\
J & \equiv &  - P_{\Lambda}
= - \alpha e^{-2 \Phi} u ^{a} \nabla_{a} \Phi
\eqnum{\ref{qldensities}b} \\
\bar{S} & \equiv &  - \bar{\Pi}_{\Phi}/\bar{N}
= - \alpha e^{-2 \Phi}(\bar{\vartheta}
+ y \bar{n}^{a} \nabla_{a} \Phi)
\eqnum{\ref{qldensities}c}\\
T & \equiv & - P_{\Phi}/\Lambda
= \alpha e^{-2 \Phi}({{\cal K}}
+ y u^{a} \nabla_{a} \Phi) \, .
\eqnum{\ref{qldensities}d}
\addtocounter{equation}{1}
\end{eqnarray}
We interpret $\bar{E}$ as the {\em quasilocal energy} 
associated with the $\bar{u}^{a}$ observers at the boundary. 
It is convenient to associate a separate energy with 
each boundary point of $\bar{\Sigma}$. For instance, 
$\bar{E}_{o} \equiv \bar{E}|_{B_{o}}$ 
is the energy associated with the 
outer boundary point. However, properly speaking, the 
full quasilocal energy associated with the 
$\bar{\Sigma}$ gravitational and dilaton fields is the 
sum 
\begin{equation}
\left.\bar{E}\right|_{B_{o}} + 
\left.\bar{E}\right|_{B_{i}} =
\left. \alpha e^{-2 \Phi}\, \bar{{{\myenn}}}^{a} \nabla_{a} \Phi
\right|_{B_{o}} - \left. \alpha e^{-2 \Phi}\, 
\bar{{{\myenn}}}^{a} \nabla_{a} \Phi
\right|_{B_{i}}\label{totalEbar}
\end{equation}
of the inner and outer boundary point contributions.
Notice that the full energy above is associated with a 
slice $\bar{\Sigma}$ which has $\bar{u}^{a}$ as its timelike 
normal {\em at the $\bar{\cal T}$ boundary}.\footnote{Technically, 
$\bar{E}_{i} + \bar{E}_{o}$ is the
energy associated with {\em any} slice in the
equivalence class determined by this condition.
The slice $\bar{\Sigma}$ can be ``wiggled" in the
interior as long as its ends remain clamped to the $\bar{\cal T}$
boundary. Similar comments are relevant for the $\Sigma$ 
energy $E_{i} + E_{o}$.} 
Also notice that, when evaluated on solutions to the field equations, 
the outer-boundary contribution to the energy $\bar{E}_{o}$,
for example, is minus the rate of change of the classical 
action (or Hamilton-Jacobi principal function) with respect
to a unit stretch in $\bar{N} |_{\bar{\cal T}_{o}}$, where 
$\bar{N} |_{\bar{\cal T}_{o}}$ controls the lapse of proper time between 
neighboring points on $\bar{\cal T}_{o}$.\cite{BY}  

The quasilocal energy associated with the $u^{a}$ observers at the 
boundary is
\begin{equation}
E \equiv \alpha e^{-2 \Phi}\, n^{a} \nabla_{a} \Phi\, ,
\end{equation}
and sum $E_{i} + E_{o}$ is the full quasilocal energy 
for a $\Sigma$ slice. The energy $E$ depends solely 
on the Cauchy data of $\Sigma$ in the same way that $\bar{E}$ 
depends solely on the Cauchy data of $\bar{\Sigma}$. 
It proves useful in the next section 
to have an energy expression for {\em each} point\footnote{Recall that 
the $\Sigma$ points are spheres, at least in the SSGR case. So we 
are not really defining a local energy, and certainly not a 
local energy {\em density}.} of $\Sigma$. However, we have a sign 
ambiguity, because each $\Sigma$ point could be viewed as an inner 
or an outer boundary point. For the sake of definiteness, define
\begin{equation}
{{\myEEE}} \equiv \alpha e^{-2\Phi} {{\myenn}}^{a} \nabla_{a} \Phi =
\alpha e^{-2\Phi} \Phi'/\Lambda\, . \label{sansE}
\end{equation}
Of course, $E = \pm {{\myEEE}}$, depending on whether it is evaluated
at the outer or the inner boundary. For some expressions, like $E^{2}$, 
the sign ambiguity cancels, so we can use $E^{2}$ or ${{\myEEE}}^{2}$.

We consider $J$ to be a {\em quasilocal momentum}. 
Notice that on-shell $J |_{t''}$, for instance, is minus 
the rate of change of the Hamilton-Jacobi principal function 
with respect to a unit stretch in $\Lambda |_{t''}$, where 
$\Lambda |_{t''}$ controls the lapse of proper distance between 
neighboring radial leaves $B$ of $t''$. Such a variation in the 
boundary data corresponds to a dilation of $t''$. Again, we
associate a $J$ with each point of a generic $\Sigma$ slice.
At a glance, one sees that $J = - P_{\Lambda}$ depends solely on 
$\Sigma$ Cauchy data.

\subsection{Boost relations and invariants}
The Eulerian observers of $\bar{\Sigma}$ at the 
$\bar{\cal T}$ boundary coincide with 
the natural observers in the boundary. We may define a set
\begin{eqnarray}
\bar{E} & = & \alpha e^{-2 \Phi}
\bar{n}^{a} \nabla_{a} \Phi \label{barobs} \eqnum{\ref{barobs}a} \\
\bar{J} & = & - \alpha
e^{-2 \Phi} \bar{u} ^{a} \nabla_{a} \Phi
\eqnum{\ref{barobs}b} \\
\bar{S} & = & - \alpha e^{-2 \Phi}
( \bar{\vartheta}
+ y \bar{n}^{a} \nabla_{a} \Phi)
\eqnum{\ref{barobs}c} \\
\bar{T} & = & \alpha e^{-2 \Phi}(\bar{{\cal K}}
+ y \bar{u}^{a} \nabla_{a} \Phi) 
\eqnum{\ref{barobs}d}
\addtocounter{equation}{1}
\end{eqnarray}
of quasilocal quantities for such observers.
Now the slice $\Sigma$ need not be clamped to the 
$\bar{\cal T}$ boundary in our formalism.
Hence, in general $\Sigma$ and $\bar{\Sigma}$ are different
slices which intersect at the same boundary point of interest.
We will refer to a switch of the spatial slice spanning a
particular boundary point as a {\em generalized boost} or
simply a boost. Properly speaking, a generalized boost is a
switch of the equivalence class of spanning slices.
The behavior 
\begin{eqnarray}
\bar{E} & = & \gamma E - v \gamma J\label{boost}
\eqnum{\ref{boost}a} \\
\bar{J} & = & \gamma J - v \gamma E
\eqnum{\ref{boost}b}\\
\bar{S} & = & \gamma S - v \gamma T +
\alpha e^{-2 \Phi}\bar{u}^{a} \nabla_{a} \eta
\eqnum{\ref{boost}c} \\
\bar{T} & = & \gamma T - v \gamma S -
\alpha e^{-2 \Phi}\bar{n}^{a} \nabla_{a} \eta
\eqnum{\ref{boost}d}
\addtocounter{equation}{1}
\end{eqnarray}
of the quasilocal quantities under boosts follows 
immediately from the boost relations (\ref{boundaryboost}) 
and the splittings (\ref{splittings}). In (\ref{boost}c,d)
the unbarred version of $\bar{S}$ is
\begin{equation}
S \equiv - \alpha e^{-2 \Phi}\left(- a_{b} n^{b}
+ y n^{b} \nabla_{b} \Phi\right)\, .
\end{equation}
Notice that $E$, $J$, $S$, and $T$ have the same dependence on 
$\Sigma$ Cauchy data that $\bar{E}$, $\bar{J}$, $\bar{S}$ and
$\bar{T}$ have on $\bar{\Sigma}$ Cauchy data. (However,
due to the appearance of acceleration terms, both $S$
and $\bar{S}$ do not depend solely on Cauchy data.)

Clearly the expression $- E^{2} + J^{2}$ is invariant
under boosts.
We may multiply it by any function of
the dilaton field or add to it any function of the
dilaton field and retain a boost-invariant expression.
Therefore, it is not completely unnatural to introduce the
invariant
\begin{equation}
M = (2\alpha\lambda)^{-1} e^{y\Phi}
\left[ - E^{2} + J^{2} + (\alpha\lambda)^{2}
\exp\left[-2y\Phi\right]\right]\, . \label{invariant}
\end{equation}
Note that $M$ has units of length for the SSGR case and units of
inverse length for 2dDG.
It turns out that on-shell (on solutions to the field equations)
$M$ is a completely conserved quantity (constant in time and space).
Moreover, for the case of SSGR we find that 
$M = M_{\scriptscriptstyle S}$,
where
$M_{\scriptscriptstyle S}$ is Kucha\v{r}'s canonical expression
for the Schwarzschild mass parameter,
\begin{equation}
M_{\scriptscriptstyle S} = \frac{P_{\Lambda}^{2}}{2R}
- \frac{R}{2}\left(\frac{R'}{\Lambda}\right)^{2} + \frac{R}{2}\, .
\end{equation}
From the four-dimensional spacetime perspective, the expression for
$M_{\scriptscriptstyle S}$ corresponds
to several mass definitions in general relativity, when spherical
symmetry is assumed. One is the Hawking mass\cite{Hawking2,SHayward}
\begin{equation}
M_{\scriptscriptstyle H} = \frac{1}{8\pi}\sqrt{\frac{A}{16\pi}}\int_{B}
{\rm d}^{2}x\sqrt{\sigma}\left[ - {\textstyle \frac{1}{2}}( k^{2}
- \ell^{2}) +{}^{2}\! R\right]\, ,
\end{equation}
where $^{2}\! R$ is the intrinsic scalar curvature of the two-surface $B$
(in our case a round sphere), and $k$ is the trace of the intrinsic
curvature of $B$ as embedded in a spanning (three-dimensional) $\Sigma$ 
slice. Likewise, $\ell$ is the trace of the extrinsic curvature
of $B$ as embedded in the timelike three-surface generated by the
integral curves of the $\Sigma$ normal, denoted
by $u^{\mu}$ for four-dimensions. 
The boost-invariant 
combination ${\textstyle \frac{1}{2}} ( k^{2} - \ell^{2})$ is more 
often written as a product of the expansions associated with the
ingoing and outgoing null normals to $B$, and the factor $A$, the 
area of $B$, ensures that the whole expression has units of
energy. For Schwarzschild $M_{\scriptscriptstyle H}$ equals
the mass parameter of the solution even for finite two-spheres. 
For general closed two-surfaces in general spacetimes,
the Hawking expression can be ``built"
as a combination of ``quasilocal boost invariants.''\cite{BYL,Lau4}

For the 2dDG case set $M_{\scriptscriptstyle W} = 2M/\alpha$.
By expressing $M_{\scriptscriptstyle W}$ in covariant form,
\begin{equation}
M_{\scriptscriptstyle W} = \lambda^{-1} e^{-2\Phi}
\left( \lambda^{2} - g^{ab} \nabla_{a}\Phi \nabla_{b}\Phi \right)\, ,
\end{equation}
we see that it is the ``local mass" of Tada and Uehara.\cite{Tada}
Such a quantity was also considered by Frolov in Ref.\cite{Frolov}.
With an argument originally given by KVK for the Schwarzschild case,
the appendix shows that $M_{\scriptscriptstyle W}$ is the canonical expression
for the
mass parameter of the Witten black hole.
The appendix also shows that the ADM energy \cite{ADM} at spatial infinity
(associated with the preferred static-time
slices) of the Witten black hole is the on-shell value of $M$. This is
the reason we would prefer to set $\alpha = 2$.

\section{Canonical theory}
This section is devoted to the canonical form of the theory. We first
sketch the Legendre transformation which yields the canonical form of the
action. We then vary the canonical action, paying strict attention to all
boundary terms. Finally, we consider the canonical transformation of
KVK and write down a new-canonical-variable version of the action
(\ref{baseaction}) which is particularly amenable to canonical
reduction.

\subsection{Form of the canonical action}
Passage to the canonical form of the action (\ref{baseaction})
demands that we write the
action in $1+1$ form as a preliminary step. This is easily done with
three ingredients. 
The first is the splitting result (\ref{splittings}b) for
the $\bar{\cal T}$ boundary extrinsic curvature 
$\bar{\vartheta}$. The second
is the identity
\begin{equation}
2y g^{ab}\nabla_{a}\Phi \nabla_{b}\Phi =
2y\left[ - \left(\frac{\dot{\Phi}}{N}\right)^{2}
+ 2 N^{r} \left(\frac{\dot{\Phi}\Phi'}{N^{2}}\right)
+ \left(\frac{\Phi'}{\Lambda\gamma}\right)^{2}\right]\, ,
\end{equation}
where $\gamma$ is the relativistic factor of the preliminary section.
The third and final ingredient is the realization that for our two
dimensional spacetime the Ricci scalar is a pure divergence,
\begin{equation}
{\cal R} = - 2 \nabla_{a}\left( {{\cal K}} u^{a} 
+ a_{b} n^{b} n^{a}\right)\, .
\end{equation}
Using these three ingredients, one can quickly cast (\ref{baseaction})
into the form
\begin{equation}
{\cal S}^{\it 1} = {\textstyle \frac{1}{4}}\,
\alpha\!\int_{\cal M} {\rm d}^{2}x
N\Lambda e^{- 2\Phi}X
+ \int_{\bar{\cal T}} {\rm d}t\,  \alpha e^{-2\Phi}
\eta \dot{\Phi}\,\, , \label{3+1action}
\end{equation}
where $X$ is the expression
\begin{eqnarray}
X & = &  - \frac{4}{N\Lambda}\left[\Lambda
{{\cal K}}\left(\dot{\Phi} - N^{r}\Phi'\right)
+ \frac{N'\Phi'}{\Lambda}\right] \nonumber \\
& & +
2y\left[ - \left(\frac{\dot{\Phi}}{N}\right)^{2}
+ 2 N^{r} \left(\frac{\dot{\Phi}
\Phi'}{N^{2}}\right) +
\left(\frac{\Phi'}{\Lambda\gamma}\right)^{2}\right]
+ 2\lambda^{2} y \exp\left[(2 - y)2\Phi\right]\, .
\end{eqnarray}
Performing the usual calculations with (\ref{3+1action}), one 
finds the following canonical action:
\begin{equation}
{\cal S}^{\it 1} = 
\int_{\cal M}{\rm d}^{2}x \left( P_{\Lambda}\dot{\Lambda} +
P_{\Phi}\dot{\Phi} - N {\cal H} - N^{r} {\cal H}_{r}\right) +
\int_{\bar{\cal T}} {\rm d}t \left( \alpha e^{-2\Phi} \eta \dot{\Phi}
- \bar{N} \bar{E}\right)\, , \label{Hamilton}
\end{equation}
where here $\bar{E}$ is short-hand for $\gamma E - v\gamma J$. It is
important to realize that in the canonical picture the equation
$\bar{E} = \alpha e^{- 2 \Phi} \bar{n}^{a} \nabla_{a} \Phi$ 
is not necessarily valid,
for equality implicitly assumes the {\em canonical equation of motion}
$P_{\Lambda} = \alpha e^{- 2\Phi} u^{a} \nabla_{a} \Phi$. 
Respectively, the Hamiltonian
constraint and the momentum constraint have the form
\begin{eqnarray}
{\cal H} & = &  \alpha^{-1} 
e^{2\Phi}\left[P_{\Phi} P_{\Lambda}
+ {\textstyle \frac{1}{2}} 
y \Lambda (P_{\Lambda})^{2}\right] \nonumber \\
& & +\,  \alpha e^{-2\Phi} \left[\left(2 - {\textstyle
\frac{1}{2}}y\right)\frac{\Phi'^{2}}{\Lambda}
- \frac{\Phi''}{\Lambda} + \frac{\Phi'\Lambda'}{\Lambda^{2}} -
{\textstyle \frac{1}{2}}\lambda^{2}\Lambda y
\exp\left[(2-y)2\Phi\right] \right] \label{constraints}
\eqnum{\ref{constraints}a}\\
{\cal H}_{r} & = & P_{\Phi} \Phi' - \Lambda P_{\Lambda}' \, .
\eqnum{\ref{constraints}b}
\addtocounter{equation}{1}
\end{eqnarray}
Notice that it is $P_{\Lambda}$ which appears differentiated in
the momentum constraint ${\cal H}_{r}$. This is to be expected,
as $\Lambda$ is a scalar density.

\subsection{Variation of the canonical action}
Straightforward manipulations establish that
the variation of the canonical
action is
\begin{eqnarray}
\delta {\cal S}^{\it 1} & = & 
\left({\rm terms\,\,which\,\,enforce\,\,
the\,\,constraints\,\,and\,\,give\,\,the}\right.
\nonumber \\
& & \left. {\rm \,\,{\em Hamiltonian}\,\,
equations\,\,of\,\,motion}\right)
+ \int^{t''}_{t'}{\rm d}r \left(P_{\Lambda}\delta\Lambda +
P_{\Phi}\delta\Phi\right) \nonumber \\
& & - \int_{\bar{\cal T}}{\rm d}t
\left[\bar{E}\delta\bar{N} + \bar{N}\bar{S}\delta\Phi
- \left(\bar{N}\bar{J} + \alpha
e^{-2\Phi}\dot{\Phi}\right)\delta\eta\right] +
\left. \alpha e^{-2\Phi} \eta
\delta\Phi\right|^{B''}_{B'}\, ,   \label{cvariation}
\end{eqnarray}
where here $\bar{E}$, $\bar{J}$, and $\bar{S}$ are
short-hand for the expressions
\begin{eqnarray}
\bar{E} & = & \gamma E - v\gamma J \label{shorthand}
\eqnum{\ref{shorthand}a} \\
\bar{J} & = & \gamma J - v\gamma E
\eqnum{\ref{shorthand}b}\\
\bar{S} & = & \gamma S - v\gamma T + \alpha
e^{-2\Phi} \bar{u}^{a} \nabla_{a} \eta\, .
\eqnum{\ref{shorthand}c}
\end{eqnarray}
As mentioned, one should be careful, for while
$E$, $J$, $S$, and $T$ are built from the
canonical variables $(\Lambda, P_{\Lambda} ; 
\Phi, P_{\Phi})$ in the same
way as before (but $S$ does not depend solely on the
canonical variables), 
{\em now the momenta $P_{\Lambda}$ and $P_{\Phi}$ need not
have the forms given in} (\ref{momenta}) 
(which are canonical equations
of motion). The term which appears multiplied 
by the variation
$\delta\eta$ vanishes when the canonical 
equation of motion
for $P_{\Lambda}$ holds. Therefore,
$\eta$ is not a quantity which is held fixed in 
our variational principle.

\subsection{Canonical Transformation}
In this subsection we perform Kucha\v{r}'s canonical transformation
on the phase-space pairs $(\Lambda, P_{\Lambda} ; \Phi, P_{\Phi})$.
In order to grasp the underlying hyperbolic geometry of this canonical
transformation, we first need to collect a few results and
observations.

Consider a black-hole solution which extremizes the
action (\ref{baseaction}) (either a Schwarzschild black hole or a
Witten black hole, depending on whether $y$ is $1$ or $2$). Associated
with this solution, there is a preferred family of static-time
slices, the collection of constant-Killing-time level surfaces.
For the Schwarzschild-black-hole case let $T(t,r)$ denote the Killing time,
and for the Witten-black-hole case let $\tau(t,r)$ denote the Killing
time. Now, given a particular $\cal M$ point $B$, we may interpret it as a
boundary point of the static-time slice $\tilde{\Sigma}$ which
contains it. Our construction defines the {\em rest frame}
$(\tilde{u}^{a} , \tilde{n}^{a})$ at $B$, where $\tilde{u}^{a}$ 
is the normal of
$\tilde{\Sigma}$ as embedded in $\cal M$ and $\tilde{n}^{a}$ is the
normal of $B$ as embedded in $\tilde{\Sigma}$. If $B$ is also
considered to be a point of the $\bar{\cal T}$ boundary, then in general
$\tilde{\Sigma}$ does not define the same frame at $B$ as
the slice $\Sigma$ or the slice $\bar{\Sigma}$
considered before. We know how to compute the energy
$\tilde{E} $ and momentum $\tilde{J}$ of the bounded
static-time slice $\tilde{\Sigma}$,
\begin{eqnarray}
\tilde{E} & = & \alpha
e^{-2\Phi}\tilde{n}^{a} \nabla_{a} \Phi
\nonumber \\
\tilde{J} & = & - \alpha
e^{-2\Phi}\tilde{u}^{a} \nabla_{a} \Phi\, .
\end{eqnarray}
Clearly $\tilde{E}$ and $\tilde{J}$ depend
on the Cauchy data of $\tilde{\Sigma}$ in the same way that $E$ and
$J$ depend on the Cauchy data of $\Sigma$.

Now, it is a fact that $\tilde{J} = 0$, which is why we refer
to  $(\tilde{u}^{a}, \tilde{n}^{a})$ as the rest frame at $B$. The
existence of the rest frame at $B$ leads 
to our key observation:
{\em at $B$ the parameter
$\varphi$ associated with the boost from the 
frame $(u^{a} , n^{a})$ defined
by $\Sigma$ to the rest frame 
$(\tilde{u}^{a} , \tilde{n}^{a})$ is
determined by the canonical
variables of $\Sigma$}.\footnote{Which need to be known 
only in a tiny neighborhood of $B$.} Indeed, with 
$w \equiv J/E$ the boost from the $\Sigma$
frame to the rest frame is parameterized by
\begin{equation}
\varphi \equiv {\textstyle \frac{1}{2}}
\log\left|\frac{1 + w}{1 - w}\right|
= {\textstyle \frac{1}{2}}
\log\left|\frac{E + J}{E - J}\right|\, .
\label{varphiboost}
\end{equation}
Notice that $\tilde{u}^{a} = \psi u^{a} + w \psi n^{a}$ with
$\psi = (1 - w^{2})^{-1/2}$. We know the expression 
(\ref{varphiboost}) is correct, because our general theory 
of boosted quasilocal energy-momentum implies that 
$E = \psi \tilde{E}$ and $J = w \psi \tilde{E}$. To get these
last relations, assume that the $\bar{\cal T}$ boundary is 
generated by the $\tilde{u}^{a}$ Eulerian history of $B$
points (identify ``tildes'' with ``bars'' and $w$
with $v$)  
and use (\ref{boost}a,b). 
At this stage we have a 
sign ambiguity in our expressions, since we did not say 
whether $B$ is an inner or an outer boundary point. 
For the sake of definiteness in what follows,
we often want to assume that $B$ is taken as an outer 
boundary point and make this distinction notationally. 
Therefore, we use ${{\myEEE}}$,
defined before in (\ref{sansE}), and 
also ${{\mydoubleu}} = J/{{\myEEE}}$
which defines ${{\myphi}} \equiv 
\frac{1}{2}\log|(1 + {{\mydoubleu}})/(1 - {{\mydoubleu}})|$.
Notice that the sign ambiguities
cancel in the expression $E \varphi 
= {{\myEEE}} {{\myphi}}$. One should be careful 
when dealing with
${{\mydoubleu}}$, $\psi$, and ${{\myphi}}$. 
Since for a classical solution the 
dilaton is a ``bad" radial coordinate in the 
sense that $\Phi' < 0$, our 
{\em unreferenced} energy ${{\myEEE}}$ is negative. The 
(in our case positive) expression for the 
quasilocal energy with the 
flat-space reference 
contribution is considered in the appendix and 
at length in Ref. \cite{BY}.
Therefore, since square roots are by 
convention positive, $\psi = 
- {{\myEEE}} /\sqrt{E^{2} - J^{2}}$.

With our canonical expression for ${{\myphi}}$ we 
straightforwardly write down a new set of constraints which
generate proper unit displacements with respect to the static-time 
slices,
\begin{eqnarray}
{\sf H} & = & \psi {\cal H} 
+ {{\mydoubleu}} \psi ({\cal H}_{r}/\Lambda)
\label{parthenon} \eqnum{\ref{parthenon}a}  \\
{\sf H}_{\vdash} & = & \psi ({\cal H}_{r}/\Lambda) +
{{\mydoubleu}} \psi {\cal H}\, . \eqnum{\ref{parthenon}b}
\addtocounter{equation}{1}
\end{eqnarray}
Since these new constraints depend only on the canonical
variables of $\Sigma$, we can consider them off the 
constraint surface in phase space.  
However, when computing a Poisson bracket $\{G, {\sf H}\}$, 
where $G$ is a functional of the canonical variables, one finds 
that all of the ``dangerous'' brackets $\{G, {{\mydoubleu}}\}$ 
which arise come multiplied by either by factor of $\cal H$ or 
${\cal H}_{r}$. 
Therefore, on-shell $\sf H$ generates proper unit displacement of the 
$\Sigma$ slice in the $\tilde{u}^{a}$ direction. 
Likewise, ${\sf H}_{\vdash}$ generates proper unit displacement of the 
$\Sigma$ slice in the $\tilde{\myenn}^{a}$ direction. 
As we show below, these generators are closely related to the 
momenta $P_{T}$ and $P_{\sf R}$ considered in KVK.

Define the new canonical variables in terms of the old ones as
follows:
\begin{eqnarray}
M & = & {\textstyle \frac{1}{2}}\,\alpha\lambda
e^{- y\Phi} (1 - F)
\label{canonical} \eqnum{\ref{canonical}a} \\
P_{M} & = & (\alpha\lambda)^{-1} e^{y\Phi} F^{-1}
\Lambda P_{\Lambda} \eqnum{\ref{canonical}b} \\
& & \nonumber \\
\Psi & = & \Phi \eqnum{\ref{canonical}c}\\
P_{\Psi} & = & P_{\Phi} + {\textstyle \frac{1}{2}}
y (1 + F^{-1})\Lambda
P_{\Lambda} + \alpha e^{-2\Phi}{{\myphi}}'\, ,
\eqnum{\ref{canonical}d}
\addtocounter{equation}{1}
\end{eqnarray}
where $M$ is the boost invariant (\ref{invariant}) and in terms
of the old variables $F$ and ${{\myphi}}$ are short-hand for
\begin{eqnarray}
F  & = & (\alpha\lambda)^{-2} \exp\left[(y - 2)2 \Phi\right]
\left[ (\alpha\Phi'/\Lambda)^{2}
- (e^{2\Phi} P_{\Lambda})^{2}\right]
\label{oldFandphi} \eqnum{\ref{oldFandphi}a} \\
& & \nonumber \\
{{\myphi}} & = & {\textstyle \frac{1}{2}}\log
\left|\frac{\alpha e^{-2\Phi}\Phi' -
\Lambda P_{\Lambda}}{\alpha e^{-2\Phi}\Phi' +
\Lambda P_{\Lambda}}\right|\,  .
\eqnum{\ref{oldFandphi}b}
\addtocounter{equation}{1}
\end{eqnarray}
Notice that ${{\myphi}} = \frac{1}{2}\log|F_{-}/F_{+}|$, where
\begin{equation}
F_{\pm} = (\alpha\lambda)^{-1} e^{y\Phi}({{\myEEE}} \mp J)\, .
\end{equation}
Evidently then, another expression for $F$ of key importance is
\begin{equation}
F = F_{+} F_{-} = (\alpha\lambda)^{-2} e^{y2\Phi}
\left( E^{2} - J^{2}\right)\, .
\label{Fexpression}
\end{equation}
For the SSGR case
$e^{-\Psi} = \lambda {\sf R}$ and $P_{\Psi} =
- {\sf R} P_{\sf R}$, in the notation of KVK and LW. Also for this case,
one finds that $P_{M} = - T'$,
because as shown in KVK the canonical expression for (minus) the radial
derivative of the Killing time is given by $- T' = (RF)^{-1}\Lambda
P_{\Lambda}$. The situation is the same for 2dDG, as in this
case $P_{M} = -\tau'$. We show in the appendix that the canonical
expression for (minus) the radial derivative of the Witten-black-hole
Killing time is $-\tau' = (\alpha\lambda)^{-1} e^{2\Phi}
F^{-1}\Lambda P_{\Lambda}$. One may prove that the transformation $(\Lambda,
P_{\Lambda} ;\Phi, P_{\Phi}) \rightarrow (M, P_{M} ; \Psi, P_{\Psi})$ is
canonical
for our boundary conditions by verifying the identity
\begin{equation}
P_{\Lambda}
\delta \Lambda + P_{\Phi} \delta \Phi
- P_{M} \delta M - P_{\Psi} \delta \Psi
= \delta\left[ \Lambda (- J + E\varphi)\right]
- \left(\alpha e^{-2\Phi} \delta \Phi {{\myphi}}\right)'\, ,
\label{generating}
\end{equation}
which upon integration over $r$ shows that the difference between the
old Liouville form and the new Liouville is an exact form. Hence, the
transformation is canonical.

Recall that for simplicity we wish to restrict our attention to
the right-static region of the relevant Kruskal diagram. Now,
in fact, for both SSGR and 2dDG $F$ is the canonical
expression for $\tilde{N}^{2}$, where $\tilde{N}$ is the lapse
function associated with the static-time slices. On-shell, the
event horizon of a particular black-hole solution is the locus of
points determined
by $F = 0$. We may ensure that we are working exclusively in a static
region of the Kruskal diagram by choosing our boundary conditions
appropriately and by excluding solutions for which $F = 0$
somewhere on $\cal M$. Where $F$ is nonzero, the above transformation
may be inverted,
\begin{eqnarray}
\Lambda & = & \left[\lambda^{-2} F^{-1}
(\Psi')^{2}\exp\left[(y - 2)2 \Psi\right] - F (P_{M})^{2}\right]^{1/2}
\label{invertedcanonical} \eqnum{\ref{invertedcanonical}a} \\
& & \nonumber \\
P_{\Lambda} & = & \frac{\alpha\lambda e^{- y \Psi}
F P_{M}}{\left[\lambda^{-2} F^{-1}
(\Psi')^{2}\exp\left[(y - 2)2 \Psi\right]
- F (P_{M})^{2}\right]^{1/2}}
\eqnum{\ref{invertedcanonical}b}\\
& & \nonumber \\
\Phi & = & \Psi \eqnum{\ref{invertedcanonical}c} \\
P_{\Phi} & = & P_{\Psi} - {\textstyle \frac{1}{2}} \alpha\lambda
y e^{- y\Psi} P_{M}(1 + F) - \alpha
e^{-2\Psi}{{\myphi}}'\, ,
\eqnum{\ref{invertedcanonical}d}
\addtocounter{equation}{1}
\end{eqnarray}
where in terms of the new variables
$F$ and ${{\myphi}}$ are short-hand for
\begin{eqnarray}
F & = & 1 - 2 (\alpha\lambda)^{-1} e^{y\Psi}M
 \label{newFandphi} \eqnum{\ref{newFandphi}a} \\
 & & \nonumber \\
{{\myphi}} & = & {\textstyle \frac{1}{2}}\log
\left|\frac{\Psi' - \lambda \exp\left[(2 - y)\Psi\right]
F P_{M}}{\Psi' + \lambda \exp\left[(2 - y)\Psi\right]
F P_{M}}\right|\,  .\eqnum{\ref{newFandphi}b}
\addtocounter{equation}{1}
\end{eqnarray}
As mentioned in the introduction,
KVK considers the canonical transformation and its inverse in all
regions of the Kruskal diagram for SSGR. Moreover, this reference
provides a detailed treatment of the (singular) behavior of the
transformation at the horizon. We expect that a similar treatment
with essentially the same results can be carried out for the 2dDG
case.

As shown in KVK for the SSGR case,
the payoff obtained by using the new variables comes when considering
the constraints (\ref{constraints}).
Since on solutions to the constraints $M$ is a
constant, one knows that $M'$ must be a sum of constraints. Indeed,
direct calculation establishes that
\begin{equation}
M' = - F^{1/2}{\sf H}\, .
\end{equation}
Since $F^{1/2}$ is the lapse $\tilde{N}$, we see that $- M'$ is the
generator of Killing-time evolution. For this reason, KVK uses the
notation $P_{T} = - M'$ (and we would likewise set $P_{\tau} =
- M'$ for the 2dDG case). Furthermore,
\begin{equation}
P_{\Psi} = - \lambda^{-1} \exp\left[(y - 2)\Phi\right]
F^{-1/2}{\sf H}_{\vdash}\,
\end{equation}
is also a sum of constraints and so weakly vanishes. KVK and LW set
$P_{\sf R} = F^{-1/2}{\sf H}_{\vdash}$. As described in KVK, the 
momenta $P_{T}$ and $P_{\sf R}$ respectively generate coordinate-scaled 
displacements of the $\Sigma$ slice along lines of constant 
${\sf R} = R$ and $T$ (or $\tau$ for 2dDG) in the Kruskal diagram.
Kucha\v{r} obtains these interpretations for $P_{T}$ and $P_{\sf R}$
via a ``parameterization at infinity'' and canonical transformations 
subsequent to the main one (\ref{canonical}). We have obtained the 
exact same interpretations via a different route. 
  
The new canonical
variables are related to the new constraints (\ref{parthenon}) 
in a very simply way: ${\sf H} = - F^{-1/2} M'$ and
${\sf H}_{\vdash} = - \lambda \exp\left[(2 - y)\Psi\right]
F^{1/2} P_{\Psi}$. Using these relations, we may write the old constraints in
terms of the new variables as
\begin{eqnarray}
{\cal H} & = & \psi {\sf H} - {{\mydoubleu}}\psi {\sf H}_{\vdash}
\nonumber \\
{\cal H}_{r} & = & \Lambda\left( \psi {\sf H}_{\vdash} - {{\mydoubleu}} \psi
{\sf H} \right)\, ,
\end{eqnarray}
where here $\Lambda$ is given in (\ref{invertedcanonical}a)
and ${{\mydoubleu}}$ must be expressed in terms of the new variables.
With the list (\ref{invertedcanonical}), it is not hard to show
that
\begin{equation}
{{\mydoubleu}}  =  - \lambda
(\Psi')^{-1}\exp\left[(2 - y)\Psi\right]  F P_{M}\, .
\end{equation}
It is now straightforward, if tedious, to
express the old constraints in terms of the new variables,
\begin{eqnarray}
{\cal H} & = & \frac{\lambda^{-1} F^{-1} \Psi'
\exp\left[(y - 2)\Psi\right] M'
+ \lambda F \exp\left[(2 - y)\Psi\right]
P_{\Psi} P_{M}}{\left[ \lambda^{-2}
F^{-1} (\Psi')^{2} \exp\left[(y - 2)2\Psi\right]
 - F (P_{M})^{2}\right]^{1/2}}
\nonumber \\
& & \\
{\cal H}_{r} & = & P_{M} M' + P_{\Psi} \Psi'\, .
\nonumber
\end{eqnarray}
As noted in KVK, in terms of the new variables it is relatively simple
to show that the Poisson bracket of $M$ with either $\cal H$ or
${\cal H}_{r}$ vanishes weakly.

\subsection{Canonical reduction}
The goal of this subsection is to use the new canonical variables
to find a reduced action principle which -in a certain sense-
corresponds to the canonical action (\ref{Hamilton}). However,
we will be adding boundary terms to (\ref{Hamilton}) before the
reduction is made, so we should clearly state what we have in mind
to begin with and why.
Several aspects of the path-integral
formulation of gravitational thermodynamics motivate fixation of
${\bar N}$ and $\Phi$ on the $\bar{\cal T}$ boundary as
the features of central importance which need to be preserved as we
modify the original action (\ref{Hamilton}). In path-integral
expressions for gravitational partition functions, the sum over
histories includes only spacetimes for which the initial and
final slices are identified. In this scenario the gauge-invariant
information of $\bar{N}$ (the lapse of proper time between the identified
initial slice $t'$ and final slice $t''$) is essentially
the inverse temperature,
which is fixed in the canonical ensemble.\cite{BMYW}
Regarding the fixation of the dilaton $\Phi$ on the
$\bar{\cal T}$ boundary, from a four-dimensional
perspective this feature allows the area of the boundary of the
system to be fixed as a boundary condition.

In what follows we modify our original canonical action
(\ref{Hamilton}) in several steps. Each step is -at least
heuristically- justified. The result will be an action
${\cal S}^{\it 1}_{\ddagger}$ (for the SSGR case this action is closely
related to one considered by LW) which is particularly amenable to
canonical reduction via the new variables. Moreover, as we will
explicitly demonstrate, the new action ${\cal S}^{\it 1}_{\ddagger}$
retains fixation of $\bar{N}$ and $\Phi$ on the $\bar{\cal T}$ boundary,
important for the above mentioned reasons, as features of its
associated variational principle. Our analysis provides some
conceptual justification for several technical steps taken in LW.

Let us go through the steps of modifying ${\cal S}^{\it 1}$. We know from
(\ref{generating}) that addition of the boundary terms
\begin{equation}
\left. - \omega[\Lambda, P_{\Lambda}, \Phi]\right|^{t''}_{t'}
= \int^{t''}_{t'} {\rm d}r \Lambda
\left(J - E\varphi\right)
\end{equation}
to the canonical action (\ref{Hamilton}) gives the new action
\begin{equation}
{\cal S}_{\dagger}^{\it 1} \equiv \int_{\cal M}{\rm d}^{2}x
\left( P_{M}\dot{M} +
P_{\Psi}\dot{\Psi} - N {\cal H}
- N^{r} {\cal H}_{r}\right) +
\int_{\bar{\cal T}} {\rm d}t
\left( \alpha e^{-2\Psi} \eta \dot{\Psi}
- \bar{N} \bar{E}\right)\, , \label{newHamilton}
\end{equation}
where here we consider all the quantities as expressed in terms
of the old variables. The vanishing of the
original set of constraints is equivalent to the vanishing of
$M'$ and $P_{\Psi}$. To take advantage of this fact,
re-express the constraint terms in the action as
$N {\cal H} + N^{r} {\cal H}_{r} = N^{M} M' +
N^{\Psi} P_{\Psi}$, where the new Lagrange multipliers are
\begin{eqnarray}
N^{M} & = & - F^{-1/2}\left(\psi N -
{{\mydoubleu}} \psi \Lambda N^{r}\right) \label{newlapseshift}
\eqnum{\ref{newlapseshift}a} \\
N^{\Psi} & = & - \lambda \exp\left[(2 - y)\Psi\right]F^{1/2}
\left(\psi
\Lambda N^{r} - {{\mydoubleu}} \psi N\right)\, . \eqnum{\ref{newlapseshift}b}
\addtocounter{equation}{1}
\end{eqnarray}
At this stage the new Lagrange multipliers still depend on the
old multipliers and the old canonical variables. In particular,
note that in terms of the old variables
\begin{equation}
N^{\Psi} = - \alpha^{-1} e^{2\Phi} \bar{N}
\left( \gamma J - v \gamma {{\myEEE}}\right)\, ,
\end{equation}
where we have used the expression (\ref{Fexpression}) and the fact that
${{\mydoubleu}} = J/{{\myEEE}}$. With the canonical equation of motion for
$P_{\Lambda}$, one can show that
$N^{\Psi} = - \alpha^{-1} e^{2\Phi} \bar{N} \bar{J} = \dot{\Phi}$.

Had we merely passed to the new canonical variables, without
redefining the Lagrange multipliers,
the variational principle associated with (\ref{newHamilton}) would have
featured fixation of $\bar{N}$ and $\Phi$ on the $\bar{\cal T}$
boundary automatically.
However, with the Lagrange multipliers
redefined (and in a way which absorbs some
of the canonical variables) all
bets are off. We must cleverly choose the
$\bar{\cal T}$ boundary term
appropriate for the new-variable version of the action.
In passing from the old constraints to $M'$ and $P_{\Psi}$, we are
effectively performing the Lorentz boost from
the frame $(u^{a}, {{\myenn}}^{a})$ to the rest
frame $(\tilde{u}^{a}, \tilde{{\myenn}}^{a})$ at each
point on $\Sigma$. A point-wise boost\footnote{Or, depending on the
viewpoint, a sphere-wise boost.}
has been performed on the old constraints and Lagrange multipliers.
However, we have not included the boundary term in
the boost. It seems that the correct way to incorporate the effect of
the boost into the boundary term is to reference the existing boost parameter
$\eta$ against the parameter $\varphi$ associated with the boost to the rest
frame. This is achieved by adding a $\bar{\cal T}$ boundary term to the action
${\cal S}^{\it 1}_{\dagger}$, with the result
\begin{equation}
{\cal S}_{\ddagger}^{\it 1} 
\equiv \int_{\cal M}{\rm d}^{2}x \left( P_{M}\dot{M} +
P_{\Psi}\dot{\Psi} - N^{M} M' - N^{\Psi} P_{\Psi}\right) +
\int_{\bar{\cal T}} {\rm d}t \left( \alpha e^{-2\Psi} \rho
\dot{\Psi} - \bar{N} \bar{E}\right)\, , \label{doubledagaction}
\end{equation}
On-shell, the boost parameter in the new action
\begin{equation}
\rho \equiv \eta - \varphi =
- {\textstyle \frac{1}{2}}\log\left|\frac{\bar{E} + \bar{J}}{\bar{E} -
\bar{J}}\right| \label{rhoparameter}
\end{equation}
is associated with the local boost between the rest frame
$(\tilde{u}^{a}, \tilde{n}^{a})$ and the boundary frame 
$(\bar{u}^{a} , \bar{n}^{a})$.
At this stage, the terms $\bar{E}$ and
$\bar{J}$ are still short-hand expressions for $\gamma E - v \gamma J$
and $\gamma J - v \gamma E$, respectively.

We now wish to express all terms in the action
${\cal S}_{\ddagger}^{\it 1}$ solely in terms
of the new variables and freely variable $N^{M}$ and $N^{\Psi}$.
Using the expressions (\ref{invertedcanonical}), one can easily
express the quasilocal energy and momentum in terms of the new
variables,
\begin{eqnarray}
E & = & \frac{\epsilon \alpha e^{-2\Psi} \Psi'}
{\left[\lambda^{-2} F^{-1}(\Psi')^{2}\exp\left[(y - 2)2 \Psi\right] - F
(P_{M})^{2}\right]^{1/2}}
\label{newEandJ} \eqnum{\ref{newEandJ}a} \\
& & \nonumber \\
J & = & \frac{- \alpha\lambda e^{- y\Psi} F P_{M}}{\left[
\lambda^{-2} F^{-1} (\Psi')^{2}\exp\left[(y - 2)2 \Psi\right] -
F (P_{M})^{2}\right]^{1/2}}\, , \eqnum{\ref{newEandJ}b}
\addtocounter{equation}{1}
\end{eqnarray}
where the factor $\epsilon \equiv (\bar{n}_{a} \bar{{{\myenn}}}^{a})$
takes care of the appropriate
sign on each of the boundary elements. Moreover, we must now regard
$N$ and $N^{r}$ (which along
with $\Lambda$ are hidden in the $v$'s and $\gamma$'s which are in turn
hidden in $\rho$) as depending
on the new variables. It is not difficult to invert the relations
(\ref{newlapseshift}) to get the needed expressions. Also, a short
calculation shows that the boundary lapse $\bar{N} = N/\gamma$ has
a fairly nice expression in terms of the new variables,
\begin{equation}
\bar{N}^{2} = F \left(N^{M}\right)^{2} - \lambda^{-2}
\exp\left[(y - 2)2 \Psi\right] F^{-1}\left(N^{\Psi}\right)^{2}\, .
\label{newNbar}
\end{equation}
However, the situation at hand remains 
quite problematic. We would
like to use the fact that $M' = 0$ to 
define a radially independent
${\bf m}(t) \equiv M(t)$ with conjugate momentum
${\bf p}(t) \equiv \int_{\Sigma} {\rm d}r P_{M}(t,r)$.
Since we have seen that $P_{M}$ is minus the radial
derivative of the Killing time, the momentum $\bf p$
would then be the difference between 
the Killing-time values for the
boundary points $B_{i}$ and $B_{o}$. 
Indeed, our canonical-reduction goal
is to insert the solutions of the constraints 
into the action to find a reduced action which is 
expressed in terms of the pair $({\bf m}, {\bf p})$.
However, $P_{M}$ appears in the $\bar{\cal T}$ boundary 
terms explicitly. Therefore, even if we perform the $r$ 
integration in the action to
define $\bf p$, the boundary
terms still contain factors of 
$P_{M}$.\footnote{The reader might suspect
that this problem was introduced when we performed 
the heuristically
justified reference 
$\eta \rightarrow \rho = (\eta - \varphi)$ to 
get the action
${\cal S}_{\ddagger}^{\it 1}$.
But the boundary term in the action 
${\cal S}_{\dagger}^{\it 1}$
suffers from the same problem. 
Indeed, $\eta$ is built from $v$
which is in turn built from $\Lambda$, 
and hence in term of the new
variables $\eta$ depends on $P_{M}$.}

A solution to the problem at hand is to make an 
appeal to the
equations of motion. For the moment, 
let us go back to considering the action
${\cal S}_{\ddagger}^{\it 1}$ as 
depending on the old variables. Notice that
{\em using the canonical equation of motion for} 
$P_{\Lambda}$, one may write
\begin{eqnarray}
\bar{J} & = & - \alpha e^{-2\Phi}
(\dot{\Phi}/\bar{N}) \label{anotherEandJ}
\eqnum{\ref{anotherEandJ}a} \\
& & \nonumber \\
\bar{E} & = &
- \epsilon \alpha e^{-2\Phi}\sqrt{\lambda^{2}
\exp\left[(2 - y)2\Phi\right] F
+ (\dot{\Phi}/\bar{N})^{2}}\, ,
\eqnum{\ref{anotherEandJ}b}
\addtocounter{equation}{1}
\end{eqnarray}
where we have appealed to the form (\ref{Fexpression}) for $F$
and again used $\epsilon$ to take care of
the appropriate sign for each boundary
element. Again, note that by convention we take the positive
square root in (\ref{anotherEandJ}b) and 
$\bar{E}$ is negative. 
It is trivial to write these new expressions 
for $\bar{E}$ and $\bar{J}$
in terms of the new variables. Fortunately, 
these expressions have no
dependence on $P_{M}$. Using these expressions 
instead of those in
(\ref{newEandJ}), we find that in terms of the new variables 
our action is again (\ref{doubledagaction}); 
but now with (i) $\bar{N}\bar{E} = - \epsilon 
\alpha e^{-2\Psi}\sqrt{\lambda^{2}
\exp\left[(2 - y)2 \Psi\right] \bar{N}^{2} F 
+ (\dot{\Psi})^{2}}$, (ii) $\bar{N}^{2}$ short-hand 
for the expression (\ref{newNbar}), (iii)
$F$ short-hand for expression (\ref{newFandphi}a), 
and (iv) the parameter specifying the boost between 
the boundary and rest frames given by
\begin{equation}
\rho = - {\textstyle \frac{1}{2}}
\log \left|\frac{\sqrt{\lambda^{2}
\exp\left[(2 - y)2 \Psi\right] \bar{N}^{2} F
+ (\dot{\Psi})^{2}}
+ \epsilon\dot{\Psi}}{\sqrt{\lambda^{2}
\exp\left[(2 - y)2 \Psi\right]
\bar{N}^{2} F +
(\dot{\Psi})^{2}}
- \epsilon\dot{\Psi}}\right|\, . \label{newrho}
\end{equation}

Now let us verify that the variational principle
associated with (\ref{doubledagaction}) and the just-given list 
(i)-(iv) does in fact possess the features 
that we demand.
The constraints and
equations of motion are
\begin{eqnarray}
M' & = & 0 \label{newEOM} \eqnum{\ref{newEOM}a} \\
P_{\Psi} & = & 0 \eqnum{\ref{newEOM}b} \\
\dot{M} & = & 0 \eqnum{\ref{newEOM}c} \\
\dot{P}_{M} & = & N^{M} \eqnum{\ref{newEOM}d} \\
\dot{\Psi} & = & N^{\Psi} \eqnum{\ref{newEOM}e} \\
\dot{P}_{\Psi} & = & 0\, . \eqnum{\ref{newEOM}f}
\addtocounter{equation}{1}
\end{eqnarray}
In terms of the old variables, we have already seen that the
equation (\ref{newEOM}e) holds when the canonical equation of
motion for $P_{\Lambda}$ is assumed.
Upon variation of the action, we find the boundary terms
\begin{eqnarray}
\lefteqn{\left(\delta 
{\cal S}_{\ddagger}^{\it 1}\right)_{\partial {\cal M}}
=} & & \nonumber \\
& &  \int^{t''}_{t'}{\rm d}r
\left(P_{M} \delta M + P_{\Psi}\delta\Psi\right)
+ \int_{\bar{\cal T}}{\rm d}t\left( \bar{\Pi}_{\bar{N}} \delta \bar{N}
+ \bar{\Pi}_{\Psi} \delta \Psi + \bar{\Pi}_{M} \delta M\right) +
\left. \alpha e^{-2 \Psi}\rho
\delta\Psi\right|^{B''}_{B'}\, , \label{deltaddagger}
\end{eqnarray}
where now the $\bar{\cal T}$ boundary momenta are the following:
\begin{eqnarray}
- \bar{E} = \bar{\Pi}_{\bar{N}} & = & \epsilon \alpha
e^{-2\Psi}\sqrt{\lambda^{2}
\exp\left[(2 - y)2 \Psi\right] F + (\dot{\Psi}/\bar{N})^{2}}
\label{dagmomenta} \eqnum{\ref{dagmomenta}a} \\
\bar{\Pi}_{\Psi} & = & {\textstyle \frac{1}{2}}
\bar{N}\left[ y\bar{E} (1 + F^{-1}) -
2 \alpha e^{-2\Psi} \bar{u}^{a} \nabla_{a}\rho\right]
\eqnum{\ref{dagmomenta}b} \\
\bar{\Pi}_{M} & = & - \epsilon N^{M}
+  (\alpha\lambda)^{-1} e^{y\Psi}
F^{-1} \bar{N}\bar{E}\, . \eqnum{\ref{dagmomenta}c}
\addtocounter{equation}{1}
\end{eqnarray}
Note that $\bar{\Pi}_{\Psi}$ is not the same as the $\bar{\Pi}_{\Phi}$
in the list (\ref{momenta}). Although the $\bar{\Pi}_{\bar{N}}$
found here is not the expression (\ref{momenta}c), it agrees with
this expression on-shell.
Recalling that $\bar{N}^{2}$ stands for (\ref{newNbar}),
one finds that $\bar{\Pi}_{M}$ vanishes when the equation of motion
(\ref{newEOM}e) holds. Therefore, $M$ need not be held fixed on
the $\bar{\cal T}$ boundary in the variational principle associated with
${\cal S}_{\ddagger}^{\it 1}$, as the equations of motion ensure that
the boundary terms with $\bar{\Pi}_{M}$ vanishes for arbitrary
variations $\delta M$ about a classical solution.

The reduced action $I^{\it 1}_{\ddagger}$, expressed in terms
of ${\bf m}$ and ${\bf p}$ defined earlier, is obtained
by solving the constraints and inserting these solutions
back into the action ${\cal S}_{\ddagger}$. From the result
(\ref{deltaddagger}) for the
variation of the action ${\cal S}_{\ddagger}^{\it 1}$, we know that
the reduced action,
\begin{eqnarray}
I_{\ddagger}^{\it 1} = \int^{t''}_{t'}{\rm d}t\, {\bf p} \dot{\bf m}
& + &   \int_{\bar{\cal T}}
{\rm d}t\, \alpha e^{-2\Psi}
\left[\epsilon\, \sqrt{\lambda^{2}
\exp\left[(2 - y)2 \Psi\right] \bar{N}^{2} F +
(\dot{\Psi})^{2}} \right. \nonumber \\
& & \left.
- {\textstyle \frac{1}{2}}\dot{\Psi}
\log \left|\frac{\sqrt{\lambda^{2}
\exp\left[(2 - y)2 \Psi\right] \bar{N}^{2} F
+ (\dot{\Psi})^{2}}
+ \epsilon \dot{\Psi}}{\sqrt{\lambda^{2}
\exp\left[(2 - y)2 \Psi\right]
\bar{N}^{2} F +
(\dot{\Psi})^{2}}
- \epsilon\dot{\Psi}}\right|\,\, \right]\, ,
\end{eqnarray}
possesses the variational principle we desire.
In the reduced action $\bar{N}$ is positive and independent, and
$F = 1 - 2 (\alpha\lambda)^{-1} e^{y\Psi} {\bf m}$. Also, in the expression
for $I^{\it 1}_{\ddagger}$ the $t'$ and $t''$ represent integration parameters
rather than spacelike slices as they did before.

\section{The thermodynamical Action}
For both the Schwarzschild and the Witten black hole
we are interested in applying the canonical
action principle to a static exterior region with spatial boundary
({\em including} the bifurcation point)
of the relevant Kruskal diagram. Such an application of the action
principle is the appropriate one 
for studying the equilibrium thermodynamics
of black holes.\cite{BMYW} In such a scenario, with the
covariant form of the action there is no inner boundary,
since the bifurcation point is a set of measure zero in the
integration over $\cal M$. Nevertheless, in the canonical picture
the bifurcation point is a boundary point of every spacelike slice,
which implies that the canonical coordinates must obey certain fall-off
conditions as the bifurcation point is approached.
Moreover, in the thermodynamical paradigm, when the initial and final
spacelike slice are identified, one must worry
about {\em regularity conditions} at the bifurcation point which
ensure that the geometry is smooth.\cite{Gibbons,BMYW,micro}
To handle these issues, we will use a technique due to
Brown and York.\cite{BYpathint} The basic idea is to work
with an inner boundary, but with boundary conditions which
effectively seal it.
The main ingredient in this technique is a new action functional,
which differs from (\ref{baseaction}) by boundary terms. The
purpose of this section is to introduce this new action principle
and to study its canonical reduction via the new canonical
variables.

\subsection{Alternative canonical action principle}

Starting with the canonical action (\ref{Hamilton}), we define
the new action
\begin{equation}
{\cal S}^{\it 1}_{*} \equiv {\cal S}^{\it 1}
- {\textstyle \frac{1}{2}}
\int_{\bar{\cal T}_{i}} {\rm d}t\bar{N}\bar{S}
+ {\textstyle \frac{1}{2}} \left. \alpha
e^{-2\Phi}\eta\right|^{B''_{i}}_{B'_{i}}\, .
\label{astaction}
\end{equation}
Note that only inner boundary and inner corner terms
have been added to the original action. 
It is easy to show that the canonical form of
${\cal S}^{\it 1}_{*}$ has the boundary terms
\begin{equation}
\left({\cal S}^{\it 1}_{*}\right)_{\partial {\cal M}} =
 \int_{\bar{\cal T}_{o}} {\rm d}t
\left[ \alpha e^{-2\Phi} \eta \dot{\Phi}
- \bar{N} \bar{E} \right] + \int_{\bar{\cal T}_{i}}
{\rm d}t \left[
{\textstyle \frac{1}{2}}\, \alpha e^{-2\Phi}\dot{\eta}
- \bar{N}\left( \bar{E} + {\textstyle \frac{1}{2}}
\bar{S}\right)\right]\, .
\end{equation}
Using the expressions (\ref{shorthand}a) and
(\ref{shorthand}c), one finds
that the inner boundary term is
\begin{equation}
\left({\cal S}^{\it 1}_{*}\right)_{\bar{\cal T}_{i}}
= - \int_{\bar{\cal T}_{i}} {\rm d}t
N\left[ E - v  J
+ {\textstyle \frac{1}{2}}( S - v T)
\right]\, . \label{astinnerbt}
\end{equation}
It is also relatively straightforward to compute
the variation of ${\cal S}^{\it 1}_{*}$.
With the result (\ref{cvariation}) it follows that
the $\bar{\cal T}$ boundary and corner contributions to
the variation of
${\cal S}^{\it 1}_{*}$  are
\begin{eqnarray}
\lefteqn{(\delta {\cal S}^{\it 1}_{*})_{\bar{\cal T}, B'', B'} =}
& & \nonumber \\
& & - \int_{\bar{\cal T}_{o}}{\rm d}t
\left[\bar{E}\delta\bar{N} +
\bar{N}\bar{S}\delta\Phi
- \left(\bar{N}\bar{J} + \alpha
e^{-2\Phi}\dot{\Phi}\right)\delta\eta\right]
+ \left. \alpha e^{-2\Phi} \eta
\delta\Phi\right|^{B''_{o}}_{B'_{o}}
\nonumber \\
& &
- \int_{\bar{\cal T}_{i}}{\rm d}t
\left[\bar{E}\delta \bar{N}
+ {\textstyle \frac{1}{2}}
e^{-2\Phi}\delta
\left(e^{2\Phi}\bar{N}\bar{S}\right)
- \left(\bar{N}\bar{J} + \alpha
e^{-2\Phi}\dot{\Phi}\right)\delta\eta\right]
+  {\textstyle \frac{1}{2}}\left. \alpha e^{-2\Phi}
\delta\eta\right|^{B''_{i}}_{B'_{i}}\,\, .
\end{eqnarray}
We describe the boundary conditions at the inner boundary
as {\em completely open} because $\bar{N}$ and
$e^{2\Phi}\bar{S}$ are held fixed, as opposed to closed
or {\em microcanonical} boundary
conditions characterized by fixation of the energy $\bar{E}$
and fixation of $\Phi$ 
(effectively the $B$ surface area).\cite{BYpathint}
With some work and the formulae in (\ref{shorthand}),
one can show that
the inner-boundary and inner-corner-point
contributions in the above
variation may be combined into the expression
\begin{equation}
(\delta {\cal S}^{\it 1}_{*})_{\bar{\cal T}_{i}, B''_{i}, B'_{i}} =
- \int_{\bar{\cal T}_{i}} {\rm d}t
\left[\left(E - v J\right)\delta N
+ {\textstyle \frac{1}{2}}
e^{-2\Phi} \delta\left(Ne^{2\Phi}S
- v Ne^{2\Phi} T\right)
- (N/\gamma^{2}) J\delta\eta\right]\, .
\label{openbc}
\end{equation}

Now we introduce fall-off conditions on the fields which seal the
inner boundary. What we have in mind is a general foliation of
our spatially
bounded static region $\cal M$. All of the spatial slices meet
at the bifurcation point, but otherwise are essentially arbitrary.
Our phase space is the set of fields $(\Lambda, P_{\Lambda};
\Phi, P_{\Phi})$ with the appropriate fall-off conditions
near the bifurcation point. The needed fall-off conditions have
already been given in LW for the specific case
of the Schwarzschild geometry.
For convenience and without loss of generality,
take the inner boundary, the bifurcation point, to be
located at $r = r_{i} = 0$
and the outer boundary to be located at $r = r_{o} = 1$.
The boundary conditions given in LW are the following:
\begin{eqnarray}
\Lambda(t,r) & = & \Lambda_{0}(t) + O(r^{2}) \label{top}
\eqnum{\ref{top}a}\\
\Phi(t,r) & = & \Phi_{0}(t) + \Phi_{2}(t) r^{2} + O(r^{4})
\eqnum{\ref{top}b} \\
P_{\Lambda}(t,r) & = & O(r^{3})
\eqnum{\ref{top}c}\\
P_{\Phi }(t,r) & = & O(r)
\eqnum{\ref{top}d}\\
N(t,r) & = & N_{1}(t)r + O(r^{3})
\eqnum{\ref{top}e}\\
N^{r}(t,r) & = & N_{1}^{r}(t)r + O(r^{3})\, ,
\eqnum{\ref{top}f}
\addtocounter{equation}{1}
\end{eqnarray}
where $O(r^{n})$ stands for a term whose magnitude as
$r \rightarrow 0$ is bounded by $r^{n}$ times a constant.
Also, as $r \rightarrow 0$, the $k$'th derivative of such a
term is similarly bounded by $r^{n-k}$ times a constant for
$1 \leq k \leq n$. Note that the time development at the inner
boundary has been arrested as the lapse vanishes there. 
One can show that
these boundary conditions are consistent with the equations of
motion. In other words, the Hamiltonian evolution preserves the above
boundary conditions, provided that the initial data obeys
both the above set of fall-off conditions
and the constraints (\ref{constraints}) on the initial spacelike
slice $\Sigma$ and provided that the lapse and shift also obey the
above fall-off conditions.\footnote{Note that
one must appeal to (\ref{Lambdanaught})
when showing that the boundary
conditions (\ref{top}) are consistent with
the $\dot{P}_{\Lambda}$ equation of motion.}
Moreover, the dynamical equation for $\Phi$ and
the above fall-off conditions imply that $\dot{\Phi}_{0} = 0$.
Imposition of the Hamiltonian constraint (\ref{constraints}a)
as $r \rightarrow 0$ yields the relation
\begin{equation}
(\Lambda_{0})^{2} = - 4 y^{-1} \lambda^{-2} \Phi_{2}
\exp\left[(y - 2)2\Phi_{0}
\right]\, , \label{Lambdanaught}
\end{equation}
which shows that $\Phi_{2}$ is negative for classical solutions.
Let us quickly compare this result with the Schwarzschild result found
in LW. Setting $R(t,r) = R_{0}(t)
+ R_{2}(t) r^{2} + O(r^{4})$ near the bifurcation point,
one finds that $\lambda R_{0} = e^{- \Phi_{0}}$
and $\lambda R_{2} = - \Phi_{2}\, e^{- \Phi_{0}}$. Therefore, for the
SSGR case the above expression is $(\Lambda_{0})^{2} = 4 R_{0} R_{2}$,
which is the LW result.

Application of the fall-off conditions (\ref{top}) to the inner boundary
term (\ref{astinnerbt}) shows that now  the new action (\ref{astaction}) has
the following form:
\begin{eqnarray}
{\cal S}^{\it 1}_{*} & = & \int_{\cal M} {\rm d}^{2}x\left(P_{\Lambda} \dot{\Lambda}
+ P_{\Phi} \dot{\Phi} - N {\cal H} - N^{r} {\cal H}_{r}\right) \nonumber \\
& & + \int_{\bar{\cal T}_{o}} {\rm d}t
\left( \alpha e^{-2\Phi}\eta \dot{\Phi}
- \bar{N} \bar{E} \right) +
{\textstyle \frac{1}{2}}\int^{t''}_{t'}
{\rm d}t  \left. \alpha e^{-2\Phi} (N'/\Lambda)\right|_{r = 0}\, ,
\label{thermoaction}
\end{eqnarray}
where in the last integral $t'$ and $t''$ now represent
integration parameters and not manifolds as they
have before.
We refer to (\ref{thermoaction}) as the {\em thermodynamical action}
because of its importance in the path-integral formulation of
gravitational thermodynamics. This is the appropriate
action with which to study the canonical ensemble for spherically
symmetric black holes.\cite{BMYW} Using (\ref{openbc}) in tandem
with the fall-off conditions (\ref{top}), one finds the following
boundary terms in the variation of the thermodynamical action:
\begin{eqnarray}
(\delta {\cal S}^{\it 1}_{*})_{\bar{\cal T}, B'', B'} & = &
{\textstyle \frac{1}{2}}\int^{t''}_{t'} {\rm d}t \left.
\alpha e^{-2\Phi}\delta\left(N'/\Lambda\right)\right|_{r = 0}
\nonumber \\
& & - \int_{\bar{\cal T}_{o}}{\rm d}t
\left[\bar{E}\delta\bar{N} +
\bar{N}\bar{S}\delta\Phi
- \left(\bar{N}\bar{J} + \alpha
e^{-2\Phi}\dot{\Phi}\right)\delta\eta\right]
+ \left. \alpha e^{-2\Phi} \eta
\delta\Phi\right|^{B''_{o}}_{B'_{o}}\,\, .
\nonumber \\
& &
\end{eqnarray}
For the $y = 1$ case this is precisely the action and variational
principle considered in LW. Notice that we have the same boundary
conditions at the outer boundary as before: $\bar{N}$ and $\Phi$ are
held fixed on this surface in the variational principle associated
with (\ref{thermoaction}). As spelled out in LW, the quantity
$N'/\Lambda$ which is fixed at the bifurcation point in the
variational principle has a direct physical interpretation. In fact,
$N'/\Lambda$ is the time rate of change of a certain boost parameter.
Each $\Sigma$ slice defines its own timelike normal $u^{a}$ at the bifurcation
point. As the $\Sigma$ slicing develops in time this vector is
continuously boosted at the rate $\left. (N'/\Lambda)\right|_{r = 0}
= N_{1}/\Lambda_{0}$.

\subsection{Canonical reduction of the thermodynamical action}
We want a new-variable version of the thermodynamical action which
is amenable to canonical reduction. From the work in $\S 4$ we already
know how to modify/handle the outer boundary term when passing to
the new canonical variables. Except for a minor difference, we will handle
the outer boundary term in the thermodynamical action just like in
the previous section. Therefore, the last section has already
addressed several of the delicate issues concerning the canonical
reduction of the thermodynamical action. However, there is a
new feature of the thermodynamical action which we need to worry about.
This new feature concerns the quantity $N'/\Lambda$ which is fixed at the
bifurcation point. We have already discussed why
fixation of $\bar{N}$ and $\Phi$ (now only at the outer boundary)
are important features of the variational principle.
The boundary integral at the bifurcation point is also of importance.
Indeed, in the thermodynamical paradigm black-hole entropy arises from
this term.\cite{BMYW,micro}
For the Schwarzschild case LW has shown that for applications
to gravitational thermodynamics is is crucial to retain fixation of
$N'/\Lambda$ at the bifurcation point when passing to a new-variable
version of the action. We also regard this feature of the action
principle as the feature of central importance which needs to be
preserved. Though essentially reviewing
the work of LW, this subsection shows how the Louko-Whiting formalism
extends to the 2dDG case. Therefore, our discussion has
relevance for the thermodynamics of pure dilaton gravity.

Let us present the quantities with which we will construct a
new-variable version of the thermodynamical action. The new canonical
variables are the same as before, since the transformation
(\ref{canonical}) remains canonical with the boundary conditions adopted
for the thermodynamical action. As before, the term
$- \left(\alpha e^{-2\Phi} \delta \Phi {{\myphi}}\right)'$ in the identity
(\ref{generating}) gives two boundary terms. The one at
the outer boundary vanishes as $\Phi$ is held fixed on this
boundary element. Moreover,
${{\myphi}}$ vanishes as $r \rightarrow 0$ so the inner boundary
term also vanishes.
Hence, upon integration (\ref{generating}) still shows that the difference
between
the old and new Liouville forms is an exact form. The new shift
$N^{\Psi}$ is again defined by (\ref{newlapseshift}b). However, following Louko
and Whiting, we define a {\em different} new lapse
\begin{equation}
{\sf N} = - N^{M} \left(\frac{yM}{\alpha}\right)
\left(\frac{\alpha\lambda}{2M}\right)^{2/y}.
\end{equation}
Recall that $M$ has units of length for the SSGR case
and units of inverse length for 2dDG. It is easy to
see that $N^{M}$ is dimensionless for both cases, and, therefore,
for both cases $\sf N$ has units of inverse length.
It turns out that this choice for the
lapse must be made in order to ensure that we retain fixation of
the boost rate $N'/\Lambda$ at
the bifurcation point as a feature of our 
variational principle. In terms of
${\sf N}$ the boundary lapse is given by
\begin{equation}
\bar{N}^{2} = \left(\frac{\alpha}{yM}\right)^{2}
\left(\frac{2M}{\alpha\lambda}\right)^{4/y}
F {\sf N}^{2} - \lambda^{-2}
\exp\left[(y - 2)2 \Psi\right] F^{-1}\left(N^{\Psi}\right)^{2}\, .
\label{differentNbar}
\end{equation}
The fall-off conditions (\ref{top}) imply the following fall-off
conditions for the new variables:
\begin{eqnarray}
M(t,r) & = & {\textstyle \frac{1}{2}}
\alpha\lambda \exp\left[- y \Psi_{0}(t)\right]
+ M_{2} (t) r^{2} + O (r^{4}) \label{newfall}
\eqnum{\ref{newfall}a} \\
\Psi(t,r) & = & \Psi_{0}(t) + \Psi_{2} r^{2} + O(r^{4})
\eqnum{\ref{newfall}b} \\
P_{M}(t,r) & = & O(r) \eqnum{\ref{newfall}c} \\
P_{\Psi}(t,r)  & = & O(r)\eqnum{\ref{newfall}d} \\
{\sf N}(t,r) & = & {\sf N}_{0}(t) + O(r^{2})\eqnum{\ref{newfall}e} \\
N^{\Psi}(t,r) & = & N_{2}^{\Psi}(t)r^{2} + O(r^{4})\, ,
\eqnum{\ref{newfall}f}
\addtocounter{equation}{1}
\end{eqnarray}
where
\begin{eqnarray}
\Psi_{0} & = & \Phi_{0} \label{newfall2} \eqnum{\ref{newfall2}a} \\
\Psi_{2} & = & \Phi_{2} \eqnum{\ref{newfall2}b} \\
M_{2} & = & - {\textstyle \frac{1}{2}} \alpha\lambda\Phi_{2}
\exp(- y\Phi_{0})\left[y + 4\lambda^{-2}(\Lambda_{0})^{-2}\Phi_{2}
\exp[(y-2)2\Phi_{0}]\right]
\eqnum{\ref{newfall2}c} \\
{\sf N}_{0} & = & - {\textstyle \frac{1}{4}} y \lambda^{2} N_{1}
\Lambda_{0}(\Phi_{2})^{-1}
\exp\left[(2 - y)2\Phi_{0}\right] \eqnum{\ref{newfall2}d} \\
N^{\Psi}_{2} & = & 2 \Phi_{2} N^{r}_{1} \eqnum{\ref{newfall2}e}\, .
\end{eqnarray}
We also have that
\begin{equation}
F = 4\lambda^{-2}(\Phi_{2})^{2}(\Lambda_{0})^{-2}\exp[(y - 2)2\Phi_{0}] r^{2}
+ O(r^{4})\, .
\end{equation}
For the SSGR case these fall-off results for the new variables
match those given in LW.
Equation (\ref{Lambdanaught}) implies that
\begin{equation}
\left. (N'/\Lambda)\right|_{r = 0} = - {\textstyle \frac{1}{4}} y \lambda^{2}
N_{1} \Lambda_{0}(\Phi_{2})^{-1} \exp\left[(2 - y)2\Phi_{0}
\right]\, ,\label{bstrate}
\end{equation}
which in turn gives ${\sf N}_{0} = N_{1}/\Lambda_{0}$.
Hence, we want to ensure that the
Lagrange multiplier ${\sf N}$ is fixed at the bifurcation point in
our new variational principle.

Let us now consider the new-variable version of the thermodynamical
action and show that it has the correct variational principle.
We could write down a general action which
covers both the SSGR and 2dDG cases, but the expression is a bit
unseemly. Therefore, let us examine both cases separately.
For the SSGR case we have
\begin{eqnarray}
{\cal S}_{\diamond}^{\it 1} & \equiv &
\int_{\cal M}{\rm d}^{2}x \left( P_{M}\dot{M}
+ P_{\sf R}\dot{\sf R} + 4{\sf N} M M'
 - N^{\sf R} P_{\sf R}\right) \nonumber \\
& &  + \left. \int_{t'}^{t''}
{\rm d}t\, 2 M^{2}{\sf N}\right|_{r = 0} + \int_{\bar{\cal T}_{o}}
{\rm d}t \left[ - \rho {\sf R}\dot{\sf R} - \bar{N}\bar{E}\right]\, ,
\label{LWaction}\end{eqnarray}
with the boundary lapse given by
\begin{equation}
\bar{N}^{2} = 16 M^{2}
F {\sf N}^{2} - F^{-1}\left(N^{\sf R}\right)^{2}\, .
\end{equation}
To get these expressions we have used the facts that $e^{-\Psi} = \lambda {\sf
R}$, $P_{\Psi} = - {\sf R} P_{\sf R}$, and $N^{\Psi} = - N^{\sf R}/{\sf R}$
(all in the notation of KVK and LW). Also, $\bar{E}$ and $\rho$ are still
given by (\ref{anotherEandJ}b) and (\ref{newrho}), respectively, but now
one must express them in terms of ${\sf R}$. This is precisely the action
considered in LW. For Witten's 2dDG model we find
\begin{eqnarray}
{\cal S}_{\diamond}^{\it 1} & \equiv &
\int_{\cal M}{\rm d}^{2}x \left[ P_{M}\dot{M}
+ P_{\Psi}\dot{\Psi} + \lambda^{-1}{\sf N} M'
 - N^{\Psi} P_{\Psi}\right] \nonumber \\
& &  \left.\int_{t'}^{t''}
{\rm d}t \lambda^{-1} M {\sf N}\right|_{r = 0} + \int_{\bar{\cal T}_{o}}
{\rm d}t \left[
\alpha e^{-2\Psi}\rho\dot{\Psi} - \bar{N}\bar{E}\right]\, ,
\label{dilatonthermo}
\end{eqnarray}
with
\begin{equation}
\bar{N}^{2} = \lambda^{-2}
F {\sf N}^{2} - \lambda^{-2} F^{-1}\left(N^{\Psi}\right)^{2}\, .
\label{CGHSWNbar}
\end{equation}
In contrast to the SSGR case, for 2dDG
the Lagrange multipler $N^{M}|_{r = 0}$ does, apart only
from a dimensionful constant, specify the boost rate of the $\Sigma$ normal
$u^{a}$ at the bifurcation point.

The variation of (\ref{LWaction})
has already been considered in LW,
so we will only consider the variation of (\ref{dilatonthermo}).
The equations of motion derived from (\ref{dilatonthermo}) are the
same as those given in (\ref{newEOM}),
except that now $\dot{P}_{M} = - \lambda^{-1} {\sf N}$.
Upon variation of the action (\ref{dilatonthermo}),
we find the boundary terms
\begin{eqnarray}
\left(\delta {\cal S}_{\diamond}^{\it 1}\right)_{\partial {\cal M}}
& = & \int^{t''}_{t'}{\rm d}r
\left(P_{M} \delta M + P_{\Psi}\delta\Psi\right)
+ \left.\int_{t'}^{t''}
{\rm d}t \lambda^{-1} M \delta {\sf N}\right|_{r = 0}\nonumber \\
& &
+ \int_{\bar{\cal T}_{o}}{\rm d}t\left( \bar{\Pi}_{\bar{N}} \delta \bar{N}
+ \bar{\Pi}_{\Psi} \delta \Psi + \bar{\Pi}_{M} \delta M\right) +
\left. \alpha e^{-2 \Psi}\rho
\delta\Psi\right|^{B_{o}''}_{B_{o}'}\, . \label{deltadiamond}
\end{eqnarray}
In the first integral $t'$ and $t''$ represent spacelike
slices, while in the second integral they are integration
parameters. The $\bar{\cal T}_{o}$ momenta in (\ref{deltadiamond})
are essentially the same as the outer-boundary ones
in (\ref{dagmomenta}), except that now $\bar{N}$ stands for
(\ref{CGHSWNbar}) and the momenta $\bar{\Pi}_{M} = \lambda^{-1} {\sf N}
+  (\alpha\lambda)^{-1} e^{y\Psi} F^{-1} \bar{N}\bar{E}$.
Like before, plugging in the explicit formula (\ref{CGHSWNbar})
for $\bar{N}^{2}$, one finds that $\bar{\Pi}_{M}$ vanishes
when the equation of motion $\dot{P}_{M} = - \lambda^{-1} {\sf N}$
holds. Therefore, $M$ need not be held fixed on $\bar{\cal T}_{o}$
in the variational principle associated with ${\cal S}_{\diamond}^{\it 1}$.
It is now straightforward to pass to the reduced form of the
thermodynamical action.

\section{Discussion}
We conclude with some comments concerning the possible
extension of the KVK and LW formalisms to other two-dimensional models
of gravity. Recently, important progress has been made
in the field of two-dimensional gravity with the realization
that a huge class of two-dimensional models can
be described within the framework of the so-called {\em
Poisson-sigma models} (PSM's) 
of Kl\"{o}sch, Schaller, and Strobl.\cite{Schaller}
For all such models there exists an absolutely conserved
quantity $C$ (referred to as a {\em Casimir function}
in the Poisson-sigma model language) which is analogous to our
$M$ expression (\ref{invariant}), and recently Kummer and Widerin
have explored the relationship between the PSM $C$
and notions of quasilocal energy for such models.\cite{WW}
Many of our results, especially those concerning our general
treatment of quasilocal energy-momentum, seem to extend
to the general PSM formalism.
In particular, the absolutely conserved quantity $C$ can
be interpreted as a quasilocal boost invariant.\cite{WL}
Extension of the canonical-reduction method of KVK to PSM
theory also seems
possible, though several technical difficulties lie in the
way. For instance, one encounters an almost limitless
variety of singularity structures when considering the
set of all PSM's.\cite{Schaller} For SSGR the canonical
transformation of KVK is singular at the horizon. Similar technical
difficulties are likely to surface when applying the KVK method
to any two-dimensional model. Since the collection of all
possible Penrose diagrams obtainable from PSM's
is so large, it is questionable whether or
not a fully unified treatment for the canonical reduction of
all PSMs is possible. On the other hand,
the richness of singularity structures in PSM gravitation offers
what is perhaps a promising testing ground for gravitational thermodynamics.
The appropriate thermodynamical action, as expressed in the LW formalism,
would be a crucial ingredient in any study of PSM thermodynamics
via reduced canonical variables.

\section{Acknowledgments}
For discussions and helpful comments I thank H. Balasin, 
J. D. Brown, W. Kummer, F. Schramm, S. Sinha, and J. W. York.
The bulk of this research has been supported by the Fonds zur
F\"{o}r\-der\-ung der wis\-sen\-schaft\-lich\-en For\-schung
(Lise Meitner Fellowship
M-00182-PHY). Also, some of this work was done
while visiting the University of North Carolina at Chapel Hill
with support provided by the National Science Foundation,
grant number PHY-9413207. Finally, this work was begun in
Pune at the Inter-University Centre for Astronomy \& Astrophysics,
with support from the Indo-US Exchange Programme and the
University Grants Commission of India.

\noindent 
{\small NOTE ADDED: 
After this project was completed, we learned of a new
paper\cite{Madhavan} by Varadarajan which also
treats the canonical reduction of two-dimensional pure
dilaton gravity in the manner of KVK. Moreover, this paper 
also considers quantization.}

\appendix

\section{Reduction (by spherical symmetry) of the Hilbert 
action with boundary terms}

Take our two-dimensional metric $g_{ab}$ from the preliminary section
and adjoin to it the metric of a round sphere. The result is the
four-dimensional spherically symmetric metric
\begin{equation}
^{4}\!g_{\mu\nu}{\rm d}x^{\mu} {\rm d}x^{\nu} =
- N^{2} {\rm d}t^{2} + \Lambda^{2} \left( {\rm d}r
+ N^{r} {\rm d}t\right)^{2} + R^{2}\left({\rm d}\theta^{2}
+ \sin^{2}\theta {\rm d}\phi^{2}\right)\, . \label{fourmetric}
\end{equation}
Now every point $B(r,t)$ of $\cal M$ is a round sphere of radius
$R(t,r)$. Therefore, our $1+1$ spacetime region $\cal M$ has now been
promoted to a four-dimensional time history ${}^{4}\!{\cal M}$ of
a three-dimensional spatial region which lies between concentric spheres.
Slices $\Sigma$ and sheets $\bar{\cal T}$ now correspond to three-dimensional
hypersurfaces of ${}^{4}\!{\cal M}$. In particular,
the boundary elements $\bar{\cal T}_{i}$ and $\bar{\cal T}_{o}$ are
now $2+1$ hypersurfaces in ${}^{4}\!{\cal M}$.

The action functional
associated with ${}^{4}\!{\cal M}$ is the standard Einstein-Hilbert 
action, complete with the $TrK$ terms needed to ensure that the
four-dimensional action principle features fixation of the
induced three-metric on all of the elements of the three-boundary
$\partial {}^{4}\!{\cal M}$.\cite{York,Hayward} In this appendix we will
insert the metric ansatz (\ref{fourmetric}) into the
four-dimensional spacetime action and then integrate out
the angular variables. This procedure yields a ``spherically
reduced'' action principle for the fields 
$g_{ab}$ and $R$ (or equivalently
$\Phi$) which are defined on the toy $1+1$ dimensional
spacetime described in the preliminary section. The variational
principle for the spherically reduced action still 
features fixation of the induced metric on the boundary, and the 
equations of motion derived from the reduced action are the 
four-dimensional Einstein equations subject to the ansatz of 
spherical symmetry; however, we do not prove this latter claim 
here. Ref. \cite{action} noted for the first time that the 
spherically reduced action reproduces the correct four-dimensional 
Einstein equations. 

The four-dimensional spacetime action is \cite{York,Hayward}
\begin{equation}
16\pi {\cal S}^{\it 1} =
\int_{{}^{4}\!{\cal M}}
{\rm d}^{4}x {\textstyle \sqrt{- {}^{4}\!g}}\Re
+ 2\int^{t''}_{t'}
{\rm d}^{3}x \sqrt{h} K
- 2\int_{\bar{\cal T}}
{\rm d}^{3}x
\sqrt{- \bar{\gamma}} \bar{\Theta}
- 2 \int^{B''}_{B'}
{\rm d}^{2}x\sqrt{\sigma}\eta\, ,\label{action}
\end{equation}
where $\Re$ is the Ricci scalar of ${}^{4}\!{\cal M}$, $h_{ij}$
is the induced three-metric on the spacelike slices $t'$
and $t''$, and $\bar{\gamma}_{ij}$ is the induced three-metric on the 
$\bar{\cal T}$ boundary elements. In (\ref{action}) we are 
using integral notations similar to those listed in 
(\ref{Tshort}). Also, $K_{\mu\nu}
\equiv - h^{\lambda}_{\mu}\, {}^{4}\nabla_{\lambda} u_{\nu}$ is the
extrinsic curvature of $t'$ or $t''$ as embedded in ${}^{4}\!{\cal M}$,
where ${}^{4}\nabla_{\mu}$ is the torsion-free covariant derivative
operator compatible with ${}^{4}\! g_{\mu\nu}$.
In this definition $u_{\mu}$ is the future-pointing normal of
$t'$ or $t''$, and we have used spacetime coordinates for convenience, so
$h_{\mu\nu} = {}^{4}\!g_{\mu\nu} + u_{\mu} u_{\nu}$. Likewise, the
extrinsic curvature of the $\bar{\cal T}$ boundary 
as embedded in ${}^{4}\!{\cal M}$
is $\bar{\Theta}_{\mu\nu} \equiv - \bar{\gamma}^{\lambda}_{\mu} \,
{}^{4}\nabla_{\lambda} \bar{n}_{\nu}$, where $\bar{n}_{\mu}$
is the outward-pointing spacelike normal of $\bar{\cal T}$
boundary and here
$\bar{\gamma}_{\mu\nu} = {}^{4}\!g_{\mu\nu} - \bar{n}_{\mu}
\bar{n}_{\nu}$. Finally, on the corners $B'$ and $B''$ (each is the
disjoint union of an inner and outer sphere) the two-metric
is $\sigma_{ab}$ and $\eta = \sinh^{-1}(u_{\mu} \bar{n}^{\mu})$.

Let us collect a few results needed for the reduction by 
spherical symmetry. The letter $\Sigma$ denotes the 
constant time slices associated with the line-element 
(\ref{fourmetric})), and the set of nonzero components 
of the $\Sigma$ extrinsic curvature tensor $K^{i}_{j}$ 
is the following:
\begin{eqnarray}
K^{r}_{r}
& = & -(N\Lambda)^{-1}\left[ \dot{\Lambda} 
- (\Lambda N^{r})'\right]
\label{nonzero} \eqnum{\ref{nonzero}a} \\
K^{\theta}_{\theta} = K^{\phi}_{\phi}
& = & - (NR)^{-1}\left( \dot{R} - N^{r} R'\right)
\eqnum{\ref{nonzero}b}
\addtocounter{equation}{1}
\end{eqnarray}
Treating the time-radial piece $g_{ab}$ of the full four-metric
(\ref{fourmetric}) as if it were a true metric in its own right,
we can compute its curvature scalar ${\cal R}[g]$. The result is
\begin{equation}
{\cal R} = - 2(N\Lambda)^{-1}
\left(\Lambda K^{r}_{r}\right)^{\bullet}
- 2(N\Lambda)^{-1}\left( \Lambda^{-1} N' -
\Lambda N^{r} K^{r}_{r}\right)'\, .
\end{equation}

Now we turn to the reduction. Let us start with the spacetime volume
integral in (\ref{action}). Consider the identity\cite{York}
\begin{equation}
\Re[{}^{4}\!g] = {}^{3}\!R[h] + K_{\mu\nu} K^{\mu\nu} -
(K)^{2} - 2\, {}^{4}\nabla_{\mu}\left(K u^{\mu} + a^{\mu}\right)\, ,
\end{equation}
where $a^{\mu}$ is the spacetime acceleration of $u^{\mu}$
and ${}^{3}\! R$ stands for the $\Sigma$ curvature scalar
(not to be confused with the radius function $R$). Use this
identity to split the volume integral in (\ref{action}) 
into three pieces:
\begin{eqnarray}
({\rm term\,\,1}) & = &
\int_{{}^{4}\!{\cal M}}{\rm d}^{4}x 
{\textstyle \sqrt{- {}^{4}\!g}}\, ^{3}\!R
\label{Hilbertterms} \eqnum{\ref{Hilbertterms}a} \\
({\rm term\,\,2}) & =&
\int_{{}^{4}\!{\cal M}}{\rm d}^{4}x
{\textstyle \sqrt{- {}^{4}\!g}}
\left[K_{ij}K^{ij} - (K)^{2}\right]
\eqnum{\ref{Hilbertterms}b} \\
({\rm term\,\,3}) & = &
- 2\int_{{}^{4}\!{\cal M}}{\rm d}^{4}x
\partial_{\mu}\left[
{\textstyle \sqrt{- {}^{4}\!g}}\left(K u^{\mu}
+ a^{\mu}\right)\right]\, .
\eqnum{\ref{Hilbertterms}c}
\addtocounter{equation}{1}
\end{eqnarray}
Focus attention on
\begin{equation}
({\rm term\,\,3}) = - 2
\int_{{}^{4}\!{\cal M}}{\rm d}^{4}x\left\{
\left[{\textstyle \sqrt{- {}^{4}\!g}}\left(K u^{t}
+ a^{t}\right)\right]^{\bullet}
+ \left[{\textstyle \sqrt{- {}^{4}\!g}}\left(K u^{r}
+ a^{r}\right)\right]'\right\}\, ,
\end{equation}
where one has
\begin{eqnarray}
u^{t} & = & 1/N \label{uanda} \eqnum{\ref{uanda}a}\\
u^{r} & = & - N^{r}/N \eqnum{\ref{uanda}b} \\
a^{t} & = & 0 \eqnum{\ref{uanda}c} \\
a^{r} & = & N^{-1}\Lambda^{-2} N'\, .
\eqnum{\ref{uanda}d}
\addtocounter{equation}{1}
\end{eqnarray}
After some work and with
${\textstyle \sqrt{- {}^{4}\!g}} = N\Lambda R^{2}\sin\theta$,
this term can be written as
\begin{eqnarray}
({\rm term\,\,3}) & = &
\int_{{}^{4}\!{\cal M}}
{\rm d}^{4}x{\textstyle \sqrt{- {}^{4}\!g}}
\left[ {\cal R} + 4K^{r}_{r} K^{\theta}_{\theta}
- 4\Lambda^{-2} (\log N)' (\log R)'\right]
\nonumber \\
& &
+ 4\int_{{}^{4}\!{\cal M}}{\rm d}^{4}x \sin\theta
\left[
\left(\Lambda R^{2} N^{r}
K^{\theta}_{\theta}\right)' - \left(\Lambda
R^{2}K^{\theta}_{\theta}\right)^{\bullet}\right]\, ,
\end{eqnarray}
where we have used $K^{\theta}_{\theta} = K^{\phi}_{\phi}$.
Straightforward manipulations show that
\begin{equation}
({\rm term\,\,2}) =
\int_{{}^{4}\!{\cal M}}{\rm d}^{4}x \sin\theta
\left[- 2K^{\theta}_{\theta}K^{\theta}_{\theta}
- 4 K^{r}_{r} K^{\theta}_{\theta}\right]\, .
\end{equation}
Recombining the three terms,
we get
\begin{eqnarray}
\int_{{}^{4}\!{\cal M}}{\rm d}^{4}x
{\textstyle \sqrt{- {}^{4}\!g}}\Re & = &
\int_{{}^{4}\!{\cal M}}{\rm d}^{4}x
{\textstyle \sqrt{- {}^{4}\!g}}\left[ {\cal R} +\, ^{3}\!R
- 2K^{\theta}_{\theta} K^{\theta}_{\theta}
- 4\Lambda^{-2} \left(\log N\right)'
\left(\log R\right)'\right] \nonumber \\
& &  + 4 \int_{{}^{4}\!{\cal M}}
{\rm d}^{4}x \sin\theta\left[
\left(\Lambda R^{2} N^{r}
K^{\theta}_{\theta}\right)' -
\left(\Lambda
R^{2}K^{\theta}_{\theta}\right)^{\bullet}\right]\, .
\end{eqnarray}
Now the explicit expression for the
$\Sigma$ Ricci scalar is
\begin{equation}
^{3}\!R = -4\Lambda^{-2} R^{-1}R''
+ 4\Lambda^{-3}R^{-1}\Lambda'R'
- 2\Lambda^{-2}R^{-2}(R')^{2} + 2R^{-2}\, .
\end{equation}
With this expression and the result for
$K^{\theta}_{\theta}$, one can show that
\begin{equation}
^{3}\!R - 2K^{\theta}_{\theta}K^{\theta}_{\theta} =
2R^{-2}\left[ g^{ab}
\nabla_{a} R \nabla_{b} R + 1
- 2\Lambda^{-1}\left(\Lambda^{-1}RR'\right)'\right]\, .
\end{equation}
Therefore, after a bit of re-shuffling, 
one finds the following expression
for the spacetime volume term in the action:
\begin{eqnarray}
\int_{{}^{4}\!{\cal M}}{\rm d}^{4}x
{\textstyle \sqrt{- {}^{4}\!g}}\Re & = &
\int_{{}^{4}\!{\cal M}}
{\rm d}^{4}x
{\textstyle \sqrt{- {}^{4}\!g}}
\left[ {\cal R} + 2 R^{-2} g^{ab}
\nabla_{a} R \nabla_{b} R + 2 R^{-2}\right]
\nonumber \\
& &  + 4\int_{{}^{4}\!{\cal M}}
{\rm d}^{4}x \sin\theta\left[
\left(\Lambda R^{2} N^{r}
K^{\theta}_{\theta}
- \Lambda^{-1}N R R'\right)' -
\left(\Lambda
R^{2}K^{\theta}_{\theta}\right)^{\bullet}\right]\, . 
\label{volume}
\end{eqnarray}

Now we turn to the boundary terms in the action. First
consider
\begin{equation}
2\int^{t''}_{t'}{\rm d}^{3}x \sqrt{h}K =
2\int^{t''}_{t'}{\rm d}^{3}x \sqrt{h} K^{r}_{r} +
4\int^{t''}_{t'}{\rm d}^{3}x
\Lambda R^{2} \sin\theta K^{\theta}_{\theta}\, . \label{K}
\end{equation}
For the $\bar{\cal T}$ boundary terms we have
\begin{equation}
- 2
\int_{\bar{\cal T}}{\rm d}^{3}x
\sqrt{-\bar{\gamma}}\bar{\Theta} =
- 2\int_{\bar{\cal T}}
{\rm d}^{3}x\sqrt{-\bar{\gamma}}
\bar{\Theta}^{t}_{t}
- 2\int_{\bar{\cal T}}
{\rm d}^{3}x\sqrt{-\bar{\gamma}}
\sigma_{ij}\bar{\Theta}^{ij}
\, ,
\end{equation}
where $\sigma_{ij}\bar{\Theta}^{ij} =
\bar{\Theta}^{\theta}_{\theta}
+ \bar{\Theta}^{\phi}_{\phi}$. Now use the
result\footnote{Proving this result is an
exercise with projection operators. Note 
that the extrinsic
curvature of $B$ as embedded in $\Sigma$ 
is defined in spacetime
coordinates by $k_{\mu\nu} 
= - \sigma_{\mu}^{\lambda}
D_{\lambda} n_{\nu}$, where $D_{\mu}$ 
is the $\Sigma$ intrinsic covariant
derivative operator compatible with 
$h_{\mu\nu}$ and $\sigma_{\mu\nu}
= {}^{4}\!g_{\mu\nu} + u_{\mu} u_{\nu} 
- n_{\mu} n_{\nu}$.
On a $\Sigma$ covector like $n_{\nu}$, 
the action of $D_{\mu}$
is $D_{\mu} n_{\nu} = h^{\lambda}_{\mu} 
h^{\kappa}_{\nu}\,\!
^{4}\!\nabla_{\lambda} n_{\kappa}$. 
Also remember that $\bar{n}^{\mu} 
= \gamma n^{\mu} + v \gamma u^{\mu}$.}
\begin{equation}
\sigma_{ij}\bar{\Theta}^{ij}
= \gamma k + v\gamma \sigma_{ij} \label{ksplit}
K ^{ij}\, .
\end{equation}
For the case at hand, $\sigma_{ij}
K ^{ij} = 2K^{\theta}_{\theta}$ and
\begin{equation}
k = \pm 2\Lambda^{-1}R^{-1}R'
\end{equation} with $-$ at the outer sphere
and $+$ at the inner sphere.
Therefore, the $\bar{\cal T}$ boundary integrals can
be expressed as
\begin{equation}
- 2\int_{\bar{\cal T}}{\rm d}^{3}x
\sqrt{-\bar{\gamma}}\bar{\Theta} =
- 2\int_{\bar{\cal T}}{\rm d}^{3}x
\sqrt{-\bar{\gamma}}
\bar{\Theta}^{t}_{t}
+ 4
\int^{\bar{\cal T}_{o}}_{\bar{\cal T}_{i}}
{\rm d}^{3}x
N \sin\theta\left[\Lambda^{-1}RR'
- {{\myvee}}R^{2}K^{\theta}_{\theta}
\right]\, . \label{Theta}
\end{equation}
To get (\ref{Theta}), we have used 
$\sqrt{-\bar{\gamma}}\gamma = NR^{2} \sin\theta$
(note that the $\gamma$'s on the left-hand side  
are different, the second is the relativistic 
factor). Also, in the above expression
\begin{equation}
\int^{\bar{\cal T}_{o}}_{\bar{\cal T}_{i}}
= \int_{\bar{\cal T}_{o}} 
- \int_{\bar{\cal T}_{i}}\, ,
\end{equation}
and ${{\myvee}} = \Lambda N^{r}/N$. 
The $v$ in (\ref{ksplit}) is
$- {{\myvee}}$ on $\bar{\cal T}_{i}$ and 
${{\myvee}}$ on $\bar{\cal T}_{o}$.

Using the results (\ref{volume}), (\ref{K}), and (\ref{Theta}),
we see that with the metric ansatz (\ref{fourmetric}) the
spacetime action (\ref{action}) reduces to
\begin{eqnarray}
16\pi {\cal S}^{\it 1} & = & \int_{{}^{4}\!{\cal M}}
{\rm d}^{4}x {\textstyle \sqrt{- {}^{4}\!g}}
\left[ {\cal R} + 2 R^{-2} g^{ab}
\nabla_{a} R \nabla_{b} R
+ 2 R^{-2}\right] \nonumber \\
& &
+ 2\int^{t''}_{t'}
{\rm d}^{3}x \sqrt{h} K^{r}_{r}
- 2\int_{\bar{\cal T}}
{\rm d}^{3}x
\sqrt{- \bar{\gamma}} \bar{\Theta}^{t}_{t}
- 2 \int^{B''}_{B'}
{\rm d}^{2}x\sqrt{\sigma}\eta\, .
\end{eqnarray} 
Now define ${{\cal K}} \equiv K^{r}_{r}$
and $\bar{\vartheta} \equiv \bar{\Theta}^{t}_{t}$.
None of the quantities in the above
action has any angular dependence. 
Therefore, we simple integrate
over the angular variables to find
\begin{eqnarray}
{\cal S}^{\it 1} & = & 
{\textstyle \frac{1}{4}}\int_{\cal M}
{\rm d}^{2}x \sqrt{-g} R^{2}
\left[ {\cal R} + 2 R^{-2} g^{ab}
\nabla_{a} R \nabla_{b} R
+ 2 R^{-2}\right] \nonumber \\
& &
+ {\textstyle \frac{1}{2}}\int^{t''}_{t'}
{\rm d}r \Lambda R^{2} {{\cal K}}
- {\textstyle \frac{1}{2}}\int_{\bar{\cal T}}
{\rm d}t \bar{N} R^{2} \bar{\vartheta}
- {\textstyle \frac{1}{2}} \left. 
R^{2} \eta\right|^{B''}_{B'}\, .
\end{eqnarray}
If one prefers, one can write the action in terms of the dilaton
$\Phi = - \log \lambda R$. The result
is precisely the action (\ref{baseaction}) from the first
section with the choices $y = 1$ and $\alpha = \lambda^{-2}$.

\section{The Witten black hole}
In this appendix we have two goals in mind. 
The first is to compute
the energy at spatial infinity (associated with the 
static-time slices $\tilde{\Sigma}$) for the Witten 
black hole. The second is to derive expressions, 
depending on the canonical variables of an arbitrary 
spacelike slice $\Sigma$, for the Witten-black-hole 
mass parameter $M_{\scriptscriptstyle W}$ and (minus) 
the radial derivative of the Killing time $- \tau'$.
Our procedure for obtaining such expressions is nearly 
identical to one found in KVK. Multiplying the 
expression we find for $M_{\scriptscriptstyle W}$
by $\alpha/2$, one finds the boost invariant $M$ given
in (\ref{invariant}).

\subsection{Line-element and asymptotic energy}
In Kruskal-type coordinates the black-hole solution of 
(vacuum) 2dDG is given by the line-element
\begin{equation}
{\rm d}s^{2} = 
- \left(M_{\scriptscriptstyle W}/\lambda - \lambda^{2}
x^{+}x^{-}\right)^{-1} {\rm d}x^{+} {\rm d}x^{-}\, ,
\label{Kruskal}
\end{equation}
along with the following expression for the dilaton:
\begin{equation}
e^{-2\Phi} = M_{\scriptscriptstyle W}/\lambda - 
\lambda^{2} x^{+}x^{-} \, .
\end{equation}
Our form of the line-element corresponds to the one 
given in Ref. \cite{Strom}, and the
discussion which follows uses the conventions of that 
reference. We are only interested in the right static region of
the Kruskal diagram associated with (\ref{Kruskal}).

We wish to compute the energy at infinity 
associated with the preferred static-time slices 
$\tilde{\Sigma}$, so we first have to
find these slices. Consider the new coordinates 
$(\tau,\sigma)$
defined by\cite{Strom}
\begin{eqnarray}
\lambda x^{-} & = & - e^{- \lambda(\tau - \sigma)}
\label{newcoordinates} \eqnum{\ref{newcoordinates}a}\\
& & \nonumber \\
\lambda x^{+} & = & e^{\lambda(\tau + \sigma)}\, .
\eqnum{\ref{newcoordinates}b}
\addtocounter{equation}{1}
\end{eqnarray}
Note that the coordinate patch $(\tau,\sigma)$ only 
covers the right static region of the Kruskal diagram. 
In terms of these coordinates
the line-element reads
\begin{equation}
{\rm d}s^{2} = \left(1 + e^{- 2\lambda\sigma} 
M_{\scriptscriptstyle W}/\lambda
\right)^{-1}
\left(- {\rm d}\tau^{2} + {\rm d}\sigma^{2}\right)\, .
\label{line}
\end{equation}
Notice that as $\sigma \rightarrow \infty$ the line 
element becomes flat. Further, notice that as $\sigma 
\rightarrow \infty$ the dilaton behaves as
\begin{equation}
\Phi = - \lambda \sigma 
- e^{-2\lambda\sigma} M_{\scriptscriptstyle W}/2\lambda
+ O(e^{-2\lambda\sigma})\, ,
\end{equation}
where $O(e^{-2\lambda\sigma})$ stands for quadratic and
higher powers in $e^{-2\lambda\sigma}$. At spatial infinity the 
black-hole solution approaches
the {\em linear dilaton}, the vacuum solution 
of the theory. For the linear
dilaton $\Phi = - \lambda \sigma$ and the line 
element is Minkowskian.

The static-time slices $\tilde{\Sigma}$ are level surfaces of
constant $\tau$. We shall pick one and evaluate the energy at an
outer boundary point $B_{o}$. Now the normal of such a point
as embedded in $\tilde{\Sigma}$ has been denoted 
$\tilde{{{\myenn}}}^{a}$. Since only the static-time slices 
are of interest now, we shall drop the tilde which appears on
$\tilde{\Sigma}$, $\tilde{{{\myenn}}}^{a}$, and other objects.
Furthermore, let us simply write $n^{a}$ for ${{\myenn}}^{a}$,
with the understanding that the rest of our analysis in this
appendix deals exclusively with an {\em outer} boundary point. 
The history of $B_{o}$ with respect to the Eulerian observers of the 
Killing time slices is the timelike sheet ${\cal T}_{o}$ defined by 
$\sigma = \sigma_{o} \geq - \infty$, where $\sigma_{o}$ is a finite 
constant. This means that the dilaton $\Phi$ is also fixed to a 
constant value $\Phi_{o}$ on ${\cal T}_{o}$. Now it turns out that 
the energy expression for $B_{o}$ as embedded in the  
static slice diverges in the limit that 
$\sigma_{o} \rightarrow \infty$.

In order to obtain a finite energy
at spatial infinity, we must reference the energy against the
linear dilaton vacuum {\em before} taking the limit. The expression
for the quasilocal energy {\em with reference point} is
\begin{equation}
E = \alpha e^{-2\Phi}\left[ n^{a} \nabla_{a} \Phi - \left.
\left( n^{a} \nabla_{a} \Phi\right)\right|^{0}\right]\,.\label{totalE}
\end{equation}
This expression has been obtain by comparison with the known expression
for quasilocal energy in general relativity,\cite{BY}
\begin{equation}
E = (8\pi)^{-1}\int_{B} {\rm d}^{2}x
\sqrt{\sigma}\left(k - k^{\it 0}\right)\, .\label{reference}
\end{equation}
In this expression $\sqrt{\sigma}$ is the square root of the
determinant of the $B$ metric (in our case that of a round 
sphere, $R^{2}\sin\theta$), $k$ is the trace of the extrinsic 
curvature of $B$ as embedded in the three-dimensional 
hypersurface $\Sigma$ of interest, and $k^{\it 0}$ is the trace 
of the extrinsic curvature of a two-surface isometric to 
$B$ which is embedded in three-dimensional flat 
Euclidean space.\footnote{Such
a construction is not possible for a generic 
two surface. Of course,
such a construction is always possible when 
$B$ is a round sphere,
the relevant case for this work.} The origin of the reference term
$k^{\it 0}$ can be traced to the freedom to add a subtraction term
(a functional of the fixed boundary data) to the four-dimensional
spacetime action without affecting the variational principle.\cite{BY}
Likewise, the reference point contribution in (\ref{totalE}) arises from the
freedom to append a subtraction term $- {\cal S}^{\it 0}$ to our base action
({\ref{baseaction}).
By inspecting (\ref{reference}),
we see that the correct way to calculate the referenced energy
is to first calculate $n^{a} \nabla_{a} \Phi - 
(n^{a} \nabla_{a} \Phi)|^{0}$ and then multiply
by the nonlinear ``determinant" factor $\alpha e^{-2\Phi}$.

We shall compute the quasilocal energy for the black hole with the
subtraction term $(n^{a} \nabla_{a} \Phi)|^{0}$ determined by the
linear-dilaton vacuum. For the black-hole solution the
outward-pointing normal to points embedded in the constant
$\tau$ slices is
\begin{equation}
n^{a} \partial/\partial x^{a} 
= \left(1 + e^{- 2\lambda\sigma} 
M_{\scriptscriptstyle W}/\lambda
\right)^{1/2}
\partial/\partial\sigma\, .
\end{equation}
With this one finds
\begin{equation}
n^{a} \nabla_{a} \Phi = - \lambda \left(1 +
e^{- 2\lambda\sigma} 
M_{\scriptscriptstyle W}/\lambda \right)^{-1/2}\, .
\end{equation}
A similar calculation for the case of the linear dilaton gives
\begin{equation}
\left. n^{a} \nabla_{a} \Phi\right|^{0} = - \lambda\, .
\end{equation}
Hence, for the point $B_{o}$ located at a $\sigma = \sigma_{o}$
as embedded in the constant-$\tau$ slice, 
the associated referenced quasilocal
energy is
\begin{equation}
E = \left.\alpha\lambda
\left[ e^{2\lambda\sigma} + M_{\scriptscriptstyle W}/\lambda\right]
\left[ 1 - \left(1 + e^{- 2\lambda\sigma}
M_{\scriptscriptstyle W}/\lambda \right)^{-1/2}\right]\right|_{\sigma =
\sigma_{o}}\, .
\end{equation}
We then have that $\lim_{\sigma_{o} \rightarrow \infty}
E = \frac{1}{2}\, \alpha M_{\scriptscriptstyle W}$.
Hence we obtain the on-shell value of $M$ given in (\ref{invariant}) as the
asymptotic energy associated with the static time slices. Note that
the asymptotic energy is $M_{\scriptscriptstyle W}$ if we make the choice
$\alpha = 2$
for 2dDG.

\subsection{Canonical expressions for $M_{\scriptscriptstyle W}$ and $-\tau'$}
Use of the dilaton itself as the radial coordinate casts
the line-element (\ref{line}) in the Schwarzschild-like form
\begin{equation}
{\rm d}s^{2} = - F {\rm d}\tau^{2} +
F^{-1} \left({\rm d}\Phi/\lambda\right)^{2}\, , \label{curvature}
\end{equation}
where $F \equiv 1 - e^{2\Phi}M_{\scriptscriptstyle W}/\lambda$. The horizon is
located at $\Phi =
- (1/2)\log(M_{\scriptscriptstyle W}/\lambda)$, or equivalently at
$R = \sqrt{M_{\scriptscriptstyle W}/\lambda^{3}}$, and we know from the Kruskal
form of the line-element
(\ref{Kruskal}) that the geometry is perfectly regular at the
horizon ($x^{+}x^{-} = 0$).
The goal now is to obtain canonical expressions for
$\tau'$ and $M_{\scriptscriptstyle W}$.

To get the desired expressions follow the method of
KVK and assume that $\tau = \tau(t,r)$ and $\Phi = \Phi(t,r)$.
It proves
convenient to define a dimensionful dilaton
$\bar{\Phi} \equiv \Phi/\lambda$. Now expand the differentials
${\rm d}\tau = \dot{\tau}{\rm d}t + \tau'
{\rm d}r$ and
${\rm d}\bar{\Phi} = \dot{\bar{\Phi}}
{\rm d}t + \bar{\Phi}' {\rm d}r$
and plug these into the line-element
(\ref{curvature}). Comparison of the result with
the ADM form of the metric (\ref{ADM})
gives the following equations:
\begin{eqnarray}
\Lambda^{2} & = & - F (\tau')^{2} + F^{-1}
(\bar{\Phi}')^{2} \label{expansion}
\eqnum{\ref{expansion}a} \\
\Lambda^{2} N^{r} & = & - F \dot{\tau}\tau'
+ F^{-1} \dot{\bar{\Phi}}\bar{\Phi}'
\eqnum{\ref{expansion}b} \\
- N^{2} + \left(\Lambda N^{r}\right)^{2}
& = & - F (\dot{\tau})^{2} + F^{-1}
(\dot{\bar{\Phi}})^{2}\, .
\eqnum{\ref{expansion}c}
\addtocounter{equation}{1}
\end{eqnarray}
 From these it is straightforward to obtain
the following expressions for the
lapse and shift:
\begin{eqnarray}
N^{r} & = & \frac{- F \dot{\tau}\tau'
+ F^{-1} \dot{\bar{\Phi}}\bar{\Phi}'}{-
F (\tau')^{2} + F^{-1}
(\bar{\Phi}')^{2}}
\label{dilatonlapseandshift}
\eqnum{\ref{dilatonlapseandshift}a} \\
& & \nonumber \\
N & = & \frac{\dot{\bar{\Phi}} \tau' -
\dot{\tau} \bar{\Phi}'}{\sqrt{- F (\tau')^{2} +
F^{-1} (\bar{\Phi}')^{2}}}\, .
\eqnum{\ref{dilatonlapseandshift}b}
\addtocounter{equation}{1}
\end{eqnarray}
In obtaining the formula for $N$, we have taken a square root.
Therefore, we need to verify that we have taken this root
in such a way that the lapse is positive in the right static region
of the Kruskal diagram,
since we want our spacelike slices to advance everywhere into the
future. Note that the dilaton is a ``bad" radial coordinate in
the sense that $\Phi \rightarrow - \infty$ as one approaches
spatial infinity, whereas the preliminary
section has assumed that the
radial coordinate $r$ increases in the direction of
spatial infinity.
Therefore, in the right static region $t = \tau$ and
$r = - \bar{\Phi}$ are ``good" coordinates,
and, using these, we see that the lapse is positive
everywhere in the right static region. The next step is to
insert the last two expressions into the formula for
the momentum
\begin{equation}
P_{\Lambda} = \alpha e^{-2\Phi} N^{-1}
(\dot{\Phi} - N^{r} \Phi')\, .
\end{equation}
After some algebra this insertion yields the
first of our desired expressions
\begin{equation}
- \tau' = (\alpha\lambda)^{-1} e^{2\Phi}
F^{-1}\Lambda P_{\Lambda}\, .
\end{equation}
Now, using this expression for $- \tau'$ in the
first equation of (\ref{expansion}),
we find the canonical expression for $F$,
\begin{equation}
F = (\alpha\lambda)^{-2}\left[(\alpha\Phi'/\Lambda)^{2} -
(e^{2\Phi} P_{\Lambda})^{2}\right] \, .
\end{equation}
Solving for $M_{\scriptscriptstyle W}$, one finds the second desired
expression,
\begin{equation}
M_{\scriptscriptstyle W} = (\alpha\lambda)^{-2} \lambda e^{2\Phi}
(P_{\Lambda})^{2} - \lambda^{-1} e^{-2\Phi} (\Phi'/\Lambda)^{2}
+ \lambda e^{-2\Phi}\, .
\end{equation}
Notice that $M = {\textstyle \frac{1}{2}} \alpha M_{\scriptscriptstyle W}$ is
precisely the boost invariant (\ref{invariant}) with the
appropriate choices for 2dDG.


\begin{thebibliography}{99}

\bibitem{Kuchar} K. V. Kucha\v{r}, Phys. Rev. {\bf D50}, 3961 (1994).

\bibitem{ADM} R. Arnowitt, S. Deser, and C. W. Misner, in
{\em Gravitation: An Introduction to Current Research}, edited
by L. Witten (Wiley, New York, 1962).

\bibitem{Thiemann} T. Thiemann and H. A. Kastrup, Nucl. Phys. {\bf B399},
211 (1993); T. Thiemann and H. A. Kastrup, Nucl. Phys. {\bf B399},
221 (1993); H. A. Kastrup and T. Thiemann, Nucl. Phys. {\bf B425},
665 (1994); T. Thiemann, Nucl. Phys. {\bf B436}, 681 (1995). 

\bibitem{LW} J. Louko and B. F.
Whiting, Phys Rev. {\bf D51}, 5583 (1995).

\bibitem{BMYW} J. W. York, Phys. Rev. {\bf D33}, 2092 (1986); H. W.
Braden, B. F. Whiting, and J. W. York, Phys. Rev. {\bf D36}, 3614
(1987); B. F. Whiting and J. W. York, Phys. Rev. Lett.  {\bf 61}, 1336
(1988); J. W. York, Physica {\bf A158}, 425 (1989); J. D. Brown, G. L.
Comer, E. A. Martinez, J. Melmed, B. F. Whiting, and J. W. York,
Class.  Quantum Grav. {\bf 7}, 1433 (1990); H. W. Braden, J. D. Brown,
B. F.  Whiting, and J. W. York, Phys. Rev. {\bf D42}, 3376 (1990);
J. W. York in {\em Conceptual Problems of Quantum Gravity}, edited by
A. Ashtekar and J. Stachel (Birkh\"{a}user, Boston, 1991);
J. D. Brown, E. A.  Martinez, and J. W. York, Phys. Rev. Lett. {\bf
66}, 2281 (1991).

\bibitem{BY} J. D. Brown and J. W. York, Phys. Rev. {\bf D47}, 1407
(1993).

\bibitem{BYL} J. D. Brown, S. R. Lau, and J. W. York, Jr.,
{\em General spacetime kinematics, the gravitational action, and quasilocal
stress-energy-momentum}, manuscript in preparation, 1996.

\bibitem{Witten} E. Witten, Phys. Rev {\bf D44}, 314 (1991).

\bibitem{CGHS} C.G. Callen, S. B. Giddings, J. A. Harvey, and A.
Strominger, Phys. Rev. {\bf D45}, R1005 (1992).

\bibitem{Hayward} G. Hayward and K. Wong, Phys. Rev. {\bf D46},
620 (1992); G. Hayward, Phys. Rev. {\bf D47}, 4778 (1993).

\bibitem{Kummer} R. B. Mann, A. Shiekh and L. Tarasov, Nucl. Phys.
{\bf B341} 134 (1990); D. Banks and M. O'Loughlin, Nucl. Phys.
{\bf B362} 649 (1991); S. D. Odintsov and I. J. Shapiro, Phys.
Lett. {\bf B263} 183 (1991), Mod. Phys. Lett. {\bf A7} 437 (1992);
I. G. Russo and A. A. Tseytlin, Nucl. Phys. {\bf B382} 259 (1992);
I. V. Volovich, Mod. Phys. Lett. {\bf A}, 1827 (1992); 
R. B. Mann, Phys. Rev.
{\bf D47} 4438 (1993); D. Louis-Martinez, J. Gegenberg,
and G. Kunstatter, Phys. Rev. {\bf D49} 5227 (1994);
J. S. Lemos and P. M. Sa, Phys. Rev.
{\bf D49} 2897 (1994); M. O. Katanaev,  W. Kummer and H. Liebl,
{\em Geometric interpretation of 
generalized dilaton gravity}, TUW-95-09,
Vienna, May 1995, gr-qc/9511009 (to appear in Phys. Rev. {\bf D}).

\bibitem{Gibbons} G. W. Gibbons and S. W. Hawking, Commun. Math.
Phys. {\bf 66}, 291 (1979).

\bibitem{action}
P. Thomi, B. Isaak, and P. H\'{a}j\'{\i}\v{c}ek, 
Phys. Rev. {\bf D30}, 1168 (1984).

\bibitem{York} J. W. York, Phys. Rev. Lett. {\bf 28}, 1082 (1972);
J. W. York, Found. Phys. {\bf 16}, 249 (1986).

\bibitem{Lau4} S. R. Lau, {\em New variables, 
the gravitational action, and boosted quasilocal 
stress-energy-momentum}, IUCAA 34/94, December 1994 and
TUW-95-18, July 1995, gr-qc/9504026 (to appear in Class.
Quantum Grav.).

\bibitem{Hawking2} S. W. Hawking, J. Math. Phys. {\bf 9}, 598 (1968).

\bibitem{SHayward} S. A. Hayward, Phys. Rev. {\bf D49}, 831 (1994).

\bibitem{Tada} T. Tada and S. Uehara, Phys. Rev. {\bf D51}, 4259 (1995);
Phys. Lett. B {\bf 305}, 23 (1993).

\bibitem{Frolov} V. P. Frolov, Phys. Rev. {\bf D46}, 5383 (1992).

\bibitem{micro} J. D. Brown and J. W. York, Phys. Rev. {\bf D47}, 1420
(1993).

\bibitem{BYpathint} J. D. Brown and J. W. York, {\em The path integral
formulation of gravitational thermodynamics}, IFP-UNC-491, CTMP/007/NCSU,
gr-qc/9405024, based on a talk presented
by J. D. Brown at the conference ``The Black Hole 25 Years After", Santiago,
Chile, January 1994.

\bibitem{Schaller}
T. Strobl, Phys. Rev. {\bf D50}, 7346 (1994);
P. Schaller and T. Strobl, Mod. Phys. Lett. {\bf A9}, 3129 (1994);
P. Schaller and T. Strobl, {\em Poisson sigma models: a generalization
of gravity-Yang-Mills systems in two dimensions}, hep-th/9411163;
P. Schaller and T. Strobl, {\em Quantization of field theories generalizing
gravtity-Yang-Mills systems on the cylinder}, gr-qc/9406027 or
{\em Lecture Notes in Physics} {\bf 436}, p. 98-122, ``Integrable Models
and Strings," eds. A. Alekseev, A. Hietamaeki, K. Huitu, A. Morozov,
and A. Niemi; T. Strobl, {\em Poisson-structure induced field theory and
models of 1+1 dimensional gravity}, Ph.D. dissertation, TU Wien, June 1994;
T. Kl\"{o}sch and T. Strobl, {\em Classical and quantum
gravity in 1+1 dimensions; part I: a unifying approach}, Technische
Universit\"{a}t Wien preprint, August 1995 (to appear in Class.
Quantum Grav.).

\bibitem{WW} W. Kummer and P. Widerin, {\em Conserved quasilocal quantities
and generally covariant theories in two dimensions}, preprint TUW-94-24,
gr-qc/9502031, Vienna, December 1994 (to appear in Phys. Rev. {\bf D}).

\bibitem{WL} W. Kummer and S. R. Lau, in preparation.

\bibitem{Madhavan} M. Varadarajan, ``Classical and quantum dynamics
of 2d vacuum dilatonic black holes," gr-qc/9508039, University of
Utah preprint, August 1995.

\bibitem{Strom} A. Strominger, {\em Les Houches lectures on black holes},
Lectures presented at the 1994 Les Houches Summer School ``Fluctuating
Geometries in Statistical Mechanics and Field Theory", gr-qc/9501071,
January 1995.

\end{thebibliography}
\end{document}